\documentclass[fleqn,usenatbib]{mnras}

\usepackage{newtxtext,newtxmath}
\usepackage[T1]{fontenc}

\DeclareRobustCommand{\VAN}[3]{#2}
\let\VANthebibliography\thebibliography
\def\thebibliography{\DeclareRobustCommand{\VAN}[3]{##3}\VANthebibliography}

\usepackage{graphicx}	% Including figure files
\title[Azimuthal temperature variations in ISO-Oph\,2]{Azimuthal temperature variations in ISO-Oph\,2 from multi-frequency ALMA observations}

\author[S. Casassus et al.]{
Simon Casassus,$^{1,2,3,4}$\thanks{E-mail: simon@das.uchile.cl}
Lucas Cieza,$^{4,5}$
Miguel C\'arcamo,$^{6,7,3,4}$
\'Alvaro Ribas, $^{8}$ 
Valentin Christiaens,$^{9}$ \newauthor
Abigali Rodríguez-Jiménez, $^{1,4}$
Carla Arce-Tord, $^{1,4}$
Trisha Bhowmik, $^{4,5,6}$ 
Prachi Chavan, $^{4,5}$  \newauthor
Camilo Gonz\'alez-Ruilova, $^{4,5}$
Rafael Mart\'inez-Brunner, $^{1,4}$
Valeria Guidotti, $^{3}$
Mauricio Leiva $^{3}$
\\
$^{1}$ Departamento de Astronom\'{\i}a, Universidad de Chile, Casilla 36-D, Santiago, Chile\\
$^{2}$ Facultad de Ingenier\'ia y Ciencias, Universidad Adolfo Ib\'a\~nez, Av. Diagonal las Torres 2640, Pe\~{n}alol\'{e}n, Santiago, Chile \\
$^{3}$ Data Observatory Foundation, Eliodoro Yáñez 2990, Providencia, Santiago, Chile\\
$^{4}$ Millennium Nucleus on Young Exoplanets and their Moons - YEMS, Chile\\
$^{5}$ N\'ucleo de Astronom\'ia, Facultad de Ingenier\'ia y Ciencias, Universidad Diego Portales, Av Ej\'ercito 441, Santiago, Chile\\
$^{6}$ University of Santiago of Chile (USACH), Faculty of Engineering, Computer Engineering Department, Chile\\
$^{7}$ Center for Interdisciplinary Research in Astrophysics and Space Exploration (CIRAS), Universidad de Santiago de Chile\\
$^{8}$ Institute of Astronomy, University of Cambridge, Madingley Road, Cambridge, CB3 0HA, UK \\
$^{9}$ Space sciences, Technologies \& Astrophysics Research (STAR) Institute, Universit\'e de Li\`ege, All\'ee du Six Ao\^ut 19c, B-4000 Sart Tilman, Belgium\\
}

\date{Accepted XXX. Received YYY; in original form ZZZ}

\pubyear{2023}

\usepackage{multirow}
\begin{document}
\label{firstpage}
\pagerange{\pageref{firstpage}--\pageref{lastpage}}
\maketitle

\begin{abstract}
  Environmental effects, such as stellar fly-bys and external
  irradiation, are thought to affect the evolution of protoplanetary
  disks in clustered star formation. Previous ALMA images at 225\,GHz
  of the ISO-Oph\,2 binary revealed a peculiar morphology in the disk
  of the primary, perhaps due to a possible fly-by with the secondary.
  Here we report on new ALMA continuum observations of this system at
  97.5\,GHz, 145\,GHz and 405\,GHz, which reveal strong morphological
  variations.  Multi-frequency positional alignment allows to interpret these
  spectral variations in terms of underlying physical
  conditions. ISO-Oph\,2A is remarkably offset from the centroid of
  its ring, at all frequencies, and the disk is lopsided, pointing at
  gravitational interactions.  However, the dust temperature also
  varies in azimuth, with two peaks whose direction connects with 
  HD\,147889, the earliest-type star in the Ophiuchus complex,
  suggesting that it is the dominant heat source.  The stellar
  environment of ISO-Oph\,2 appears to drive both its density
  structure and its thermal balance.
\end{abstract}

\begin{keywords}
protoplanetary discs -- stars: individual: PDS\,70, ISO-Oph\,2 --  techniques: interferometric -- stars: pre-main-sequence

\end{keywords}

%%%%%%%%%%%%%%%%%%%%%%%%%%%%%%%%%%%%%%%%%%%%%%%%%%

%%%%%%%%%%%%%%%%% BODY OF PAPER %%%%%%%%%%%%%%%%%%

\section{Introduction}

%***Para on single disc evolution and current surveys***

%***Para on how the environment may affect that picture***

%***Para on ISO-Oph\,2 and previous work***

%***Para on multi-freq ODISEA****

Interferometric observations at (sub)-millimeter wavelengths can
resolve circumstellar disks at subarcsecond resolution and trace the
thermal continuum emission due to dust
\citep[e.g. ][]{Andrews2009ApJ...700.1502A,
  Andrews2010ApJ...723.1241A, Isella2009ApJ...701..260I}. Also,
multi-wavelength (sub)-millimeter data can help constrain dust
properties such as the maximum grain size
\citep[][]{Guilloteau2011A&A...529A.105G}.  With unprecedented
capabilities, the Atacama Large Millimetre/sub-millimeter Array (ALMA)
has revolutionized the field over the last decade. ALMA has already
surveyed most star-forming regions in nearby (distances $d < 300$\,pc)
molecular clouds, including Chameleon
\citep{Pascucci2016ApJ...831..125P, Villenave2021A&A...653A..46V},
Lupus \citep{Ansdell2016ApJ...828...46A, Ansdell2018ApJ...859...21A},
Taurus \citep{Long2018ApJ...869...17L,Long2019ApJ...882...49L}, and
Ophiuchus \citep{Cox2017ApJ...851...83C,Cieza2019MNRAS.482..698C}.
Even though these surveys have mostly been carried out in a single
frequency at a modest resolution (0\farcs1 - 0\farcs2), they still
allow us to investigate disk properties as a function of different
variables, such as IR Class \citep{Williams2019ApJ...875L...9W},
stellar mass
\citep{Barenfeld2016ApJ...827..142B,Pascucci2016ApJ...831..125P}, age
\citep{Ansdell2018ApJ...859...21A,Ruiz-Rodriguez2018MNRAS.478.3674R},
and stellar-multiplicity
\citep{Cox2017ApJ...851...83C,Zurlo2020MNRAS.496.5089Z,Zurlo2021MNRAS.501.2305Z}.

In models of clustered star formation, the stellar environment affects
disk structure and evolution
\citep[e.g.][]{Haworth2021MNRAS.503.4172H, WinterHaworth2022EPJP..137.1132W,
  Wilhelm2023MNRAS.520.5331W}, both through external irradiation,
which may lead to photo-evaporation, and through gravitational
interaction, including disk truncation and accretion bursts. The Orion
proplyds \citep[][]{ODell1994ApJ...436..194O} are spectacular examples
of the impact of environment through external
photo-evaporation. Demographic surveys in the Orion nebula cluster
show that disk structure is determined in part by the distance to
$\theta^1$\,Ori\,C
\citep[][]{Mann2014ApJ...784...82M,Eisner2018ApJ...860...77E}. In
turn, models of flybys may explain the structures seen in several
binary disks
\citep[e.g.][]{Dong2022NatAs...6..331D,Cuello2023EPJP..138...11C}. However,
the probability for witnessing such close encounters (with a crossing
time of $\lesssim 400$\,yr within 500\,au at a typical relative
velocity of $\sim$6\,km\,s$^{-1}$) is very small compared to the disk
lifetime ($\sim$10\,Myr), and isolated spiral systems haven been shown
not to have undergone recent stellar encounters \citep[][ in the past
$10^{4}$\,yr]{Shuai2022ApJS..263...31S}. In any case, whichever the
mechanism, the environment bears a significant role in exoplanet
demographics
\citep[][]{Winter2020Natur.586..528W,Longmore2021ApJ...911L..16L}.

The demographic surveys also allow us to identify particularly
interesting targets for follow-up studies. Such is the case for the
ISO-Oph 2 system, a wide separation (240 au) binary targeted by the
Ophiuchus DIsc Survey Employing ALMA \citep[ODISEA,
][]{Cieza2019MNRAS.482..698C} in band-6 (230\,GHz).  ISO-Oph\,2 was
observed at 0\farcs02 resolution, also in band-6, as part of the
high-resolution follow-up of the brightest ODISEA targets
\citep[][]{Cieza2021MNRAS.501.2934C}. The high-resolution observation
showed that the disk around the primary consists of two
non-axisymmetric rings and that the disk around the secondary is a
narrow ring with a 2\,au inner radius and a 1\,au width
\citep[][]{Gonzalez-Ruilova2020ApJ...902L..33G}.  Furthermore, the
$^{12}$CO data show a bridge of gas connecting both disks, suggesting
that the binary is interacting, and is perhaps in a flyby orbit.

Another particularly interesting aspect of ISO-Oph\,2 is that, among
the  long-baseline ODISEA sample \citep[][]{Cieza2021MNRAS.501.2934C}, it is the closest to HD\,147889 \citep[B2IV,
B3IV, ][]{Casassus2008}, the earliest-type star in the Ophiuchus
complex. This proximity raises a question on the role of external
irradiation in the thermal balance in the outer ring of ISO-Oph\,2.
ISO-Oph\,2 is thus an interesting case-study for the effect of  the
stellar environment on protoplanetary disk evolution, both in
gravitational interactions and external irradiation.

The ODISEA project has recently been extended to multi-frequency
observations covering over 90 objects in ALMA Band-4 (at 145\,GHz,
Chavan et al. in prep) and Band-8 (405\,GHz, Bhowmik et al. in prep.;
Cieza et al. in prep) in order to better constrain the physical
properties of the Ophiuchus disks when combined with existing data. A
crucial aspect of such an analysis is the alignment of the
multi-frequency data, which might not be acquired with the same phase
center, or could be affected by pointing errors, that could bias
spectral trends such as spectral index maps.
\citet[][]{CasassusCarcamo2022MNRAS.513.5790C} proposed a strategy for
the alignment of multi-epoch and multi-configuration
radio-interferometric data, although their application was restricted
to the same correlator setups. It is interesting to investigate whether  the same
strategy might be applied to multi-frequency data.

A crucial aspect of image synthesis is the process of image
restoration, which conveys imaging residuals in the final images,
along with a well-defined angular resolution. In the last couple of
years a technique, usually referred to as the ``JvM correction'', has
recently been incorporated in image restoration
\citep[][]{JvM1995AJ....110.2037J, Czekala2021ApJS..257....2C}. The
JvM correction reduces the noise and residuals in the final
images. Here we stress that the present analysis does not make us of
this technique, as the resulting improvement in dynamic range is
spurious. The proof, provided by
\citet[][]{CasassusCarcamo2022MNRAS.513.5790C}, may not have been
clear enough since applications of the JvM correction have become
widespread. Here we attempt to clarify some of the aspects of the
proof in Appendix\,\ref{sec:JvM}.

This article reports on a multi-frequency analysis of ISO-Oph\,2. The new
observations, along with our alignment strategy, are described in
Sec.\,\ref{sec:obs}. The data show strong morphological variations
with frequency, which we interpret in terms of underlying physical
conditions in Sec.\,\ref{sec:analysis}. We conclude, in
Sec.\,\ref{sec:conc}, on a particularly strong impact of the
environment in the case of ISO-Oph\,2.

\section{Observations} \label{sec:obs}

\subsection{Data acquisition}

The Band\,6 ALMA observations of ISO-Oph\,2 are described in
\citet{Cieza2021MNRAS.501.2934C} and
\citet{Gonzalez-Ruilova2020ApJ...902L..33G}. The new ALMA
observations, in Band\,3, Band\,4 and Band\,8, were acquired as part
of ALMA programmes {\tt 2019.1.01111.S}, {\tt 2021.1.00378.S} and {\tt
  2022.1.01734.S}. An observation log can be found in
Table\,\ref{table:log}, and includes a nomenclature for the data-sets.

\begin{table*}
  \caption{Observation log. All dataset are original except for 225\,GHz, i.e. all scheduling blocks for B6, which have previously been reported in 
\citet{Cieza2021MNRAS.501.2934C} and \citet{Gonzalez-Ruilova2020ApJ...902L..33G} (but with different synthesis imaging tools)}
 \label{table:log}
 \begin{tabular}{cccccccccc}
  \hline
    $\nu^{\rm a}$  & Date & $\Delta t^{\rm b}$       & Baseline   & pwv$^{\rm c}$ & \multicolumn{2}{c}{Dataset} & \multicolumn{3}{c}{Beam \& Noise$^{\rm d}$}   \\
      &    &   & Range (m) &                &  \multicolumn{2}{c}{Code} &r=0  & r=1 & r=2                        \\
  \hline
   \multirow{2}{*}{405} &04-Aug-2022 & 56.4\,s  & 15 - 1300 & 0.6  &  & \multirow{2}{*}{B8}  &
                                                                                                         \multirow{2}{*}{0.157$\times$0.132/88 \& 340} & \multirow{2}{*}{0.195$\times$0.164/84 \& 260}   &    \multirow{2}{*}{0.209 $\times$ 0.173 / 83 \& 270}  \\
                  &  11-Aug-2022 & 56.4\,s     & 15 - 1300    & 0.5  &       &    &    &    &              \\  \hline   % [5pt]
   \multirow{5}{*}{225} & 12-Jun-2019 & 15\,mn  & 83 - 16196 & 1.2  & \multirow{2}{*}{LB19} &\multirow{5}{*}{B6}  & \multirow{5}{*}{0.028$\times$0.017 / -32  \& 20} & \multirow{5}{*}{0.034$\times$0.025 / -23  \& 11}   &    \multirow{5}{*}{0.036$\times$ 0.027 / -19  \& 10}  \\ 
 & 21-Jun-2019 & 24\,mn  & 83 - 16196 & 0.9  &  &  &  & \\
 & 13-Jul-2017 & 20\,s  & 16 - 2647 & 1.95  & \multirow{3}{*}{SB17} &  &  &\\
 & 13-Jul-2017 & 20\,s  & 16 - 2647 & 1.8  &  &  &  &\\
 & 14-Jul-2017 & 20\,s  & 16 - 2647 & 1.1  &  &  &  &  \\ \hline   % [5pt]
   \multirow{3}{*}{145} &20-Jul-2022 & 24\,s  & 15 - 2617 & 3.0  &  & \multirow{3}{*}{B4}  &
                                                                                                         \multirow{3}{*}{0.228 $\times$ 0.161 / 77  \& 100} & \multirow{3}{*}{0.346 $\times$ 0.237 / 79     \& 67}   &    \multirow{3}{*}{0.385 $\times$ 0.266 / 79 \& 67}  \\
                  &  21-Jul-2022 & 24\,s     & 15 - 2617    & 2.4  &       &    &    &    &              \\ 
                   &  21-Jul-2022 & 24\,s     & 15 - 2617    & 2.7  &       &    &    &    &              \\ \hline
   \multirow{3}{*}{97.5} &27-Jul-2021 & 128\,s  & 15 - 3321 & 0.6  &  & \multirow{3}{*}{B3}  &
                                                                                                         \multirow{3}{*}{0.157 $\times$ 0.108 / 88  \& 38} & \multirow{3}{*}{0.260 $\times$ 0.199 / -86     \& 24}   &    \multirow{3}{*}{0.288 $\times$ 0.224 /-83 \& 23}  \\
                  &  31-Oct-2021 & 128\,s     & 63 - 6855    & 1.0  &       &    &    &    &              \\ 
                   &  03-Nov-2021 & 128\,s     & 47 - 5185    & 1.4  &       &    &    &    &              \\ \hline
  
 \end{tabular}\\
 $^{\rm a}$~center frequency in GHz
 $^{\rm b}$~time on-source 
 $^{\rm c}$~column of precipitable water vapour, in mm 
 $^{\rm d}$~the beam  major axis (bmaj, arcsec), minor axis (bmin, arcsec) and direction
    (bpa, degrees) and noise (rms, $\mu$Jy\,beam$^{-1}$) are reported in the format bmaj$\times$bmin/bpa\& rms, for a choice of 3 Briggs robustness parameter $r$.
\end{table*}

\subsection{Imaging, self-calibration and alignment}

Automatic self-calibration was performed with the {\sc OOselfcal}
package, described in \citet[][]{CasassusCarcamo2022MNRAS.513.5790C},
which we re-baptised to ``Self-calibratioN Object-oriented
frameWork'', i.e. {\sc snow}\footnote{see Data Availability}. In
brief, {\sc snow} applies the self-calibration tasks {\tt gaincal} and
{\tt applycal} from the CASA package, but replaces the imager {\tt
  tclean} by {\sc uvmem} \citep[][]{Casassus2006,
  Carcamo2018A&C....22...16C}. Here {\sc uvmem} was run with
pure-$\chi^2$ optimization, i.e. without any regularization other than
the requirement of image positivity.  Image restoration was performed
with natural weights (Briggs robustness parameter $r=2$) for the
self-calibration iterations, and with various choices of weights for
the final images.

Each individual scheduling block, corresponding to the rows in
Table\,\ref{table:log}, were self-calibrated individually before
concatenation. Significant improvements were obtained only for B8,
where the peak signal-to-noise ratio (PSNR) increased from 29 and 36
to 79 and 74 after 4 rounds of phase-only calibration (with solution
intervals set to the scan length, 64\,s, 32\,s and 15\,s) and one
round of amplitude and phase calibration (for the scan length). We
aligned both scheduling blocks in B8 with the {\sc VisAlign} package
\citep[][]{CasassusCarcamo2022MNRAS.513.5790C}, but with the
corrections described in the Appendix\,\ref{sec:figofmerit}. Choosing 04-Aug-2022 as the
reference, the corresponding astrometric shift is
$\Delta x = -100\pm3$\,mas in the direction of R.A. and
$\Delta y = -31\pm3$\,mas in Dec., while the flux scale correction
factor is $\alpha_R = 1.20\pm0.01$. We note the very large astrometric
shift, of about half a beam (in natural weights). Such a shift is
larger than the nominal pointing accuracy of $1/10$ the clean beam,
and may reflect poor weather in either of the two epochs (the same
procedure applied before self-calibration yields an even larger shift,
$\alpha_R = 1.20\pm0.02$, $\Delta_x= -99\pm 6$\,mas and
$\Delta_y= -53\pm 5\,$mas). For B8, the concatenated scheduling blocks
have PSNR of 100, with no further improvement for phase-only
calibration, and but a small increase to 102 after amplitude and phase
self-calibration.

The resulting continuum dataset was aligned to our choice of
astrometric reference, which is the B6 dataset. We applied {\sc
  VisAlign} without scaling in flux. B6 and B8 have very different
phase centers, which may lead to the propagation of large numerical
errors when performing the alignment in the $uv$-plane (as with {\sc
  VisAlign}). We therefore performed the alignment in two steps. First
we applied a coarse shift, corresponding to the difference between the
nominal phase centers, or $\Delta x = -149.7$\,mas and
$\Delta y = -57$\,mas. We then optimized the small shift, which
yielded $\alpha_R = 0.240 \pm 0.003$, $\Delta x = 179\pm2$\,mas and
$\Delta y = 50\pm2$\,mas. We stress that, in this application of {\sc
  VisAlign}, across different ALMA bands, we set $\alpha_R=1$.

For our astrometric reference
dataset, B6, the coarse shift in the alignment of SB17 to LB19 
was $\Delta x = -16$\,mas and
$\Delta y = 82$\,mas. The optimization of the residual shift yielded
$\alpha_R = 1.02 \pm 0.06$, $\Delta x = 13\pm9$\,mas and
$\Delta y = -56\pm8$\,mas.  Self-calibration did not yield any
improvement for B6, and the imaging residuals are thermal.

For B4, self-calibration yielded a small improvement in PSNR for all
concatenated scheduling blocks, from 60 to 65 after amplitude and
phase calibration. However, the aligment of each scheduling block, for
which we chose 21-Jul as the reference, revealed an intriguing anomaly
in flux. The shift of 20-Jul to 21-Jul was $\alpha_R = 1.00 \pm 0.02$,
$\Delta x = -50\pm9$\,mas and $\Delta y = 63\pm8$\,mas. However, the
shift of the second 21-Jul block to the first, which were observed
consecutively, yielded $\alpha_R = 1.05 \pm 0.02$,
$\Delta x = -51\pm9$\,mas and $\Delta y = 22\pm6$\,mas, corresponding
to a 5\% flux scale difference. Although barely at $3\sigma$, this
flux scale difference is exactly as obtained when comparing the total
flux densities. Such a flux scale difference is still well within the
absolute flux calibration accuracy of ALMA, considered to be of
$\sim 5\%$ in Band\,4
\citep[e.g.][]{ALMA_technical_handbook2019athb.rept.....R}.

The self-calibration procedure for B3 improved PSNR from 28 to 30 for
27-Jul-2021, after 3 rounds of phase-only self-calibration (with
solution intervals set to the scan length, 64\,s, 32\,s and 15\,s),
and very small improvements from 20 to 21 for 31-Oct-2021 and from 26
to 28 for 03-Nov-2021, both after a single round of phase-calibration.
We chose 27-Jul-2021 as the reference, and shifted 31-Oct-2023 by
$\alpha_R = 1.03\pm0.03$, $\Delta_x= -11\pm7 \,$mas and
$\Delta_y = 35\pm6\,$mas, and 03-Nov-2021 by $\alpha_R= 0.98\pm0.02$,
$\Delta_x= 38\pm 7$\,mas and $\Delta_y= 15\pm 7 $\,mas. The
concatenated dataset reaches a PNSR of 44 in natural weights, with no
further improvements with self-calibration.

\begin{figure}
  \centering
  \includegraphics[width=\columnwidth,height=!]{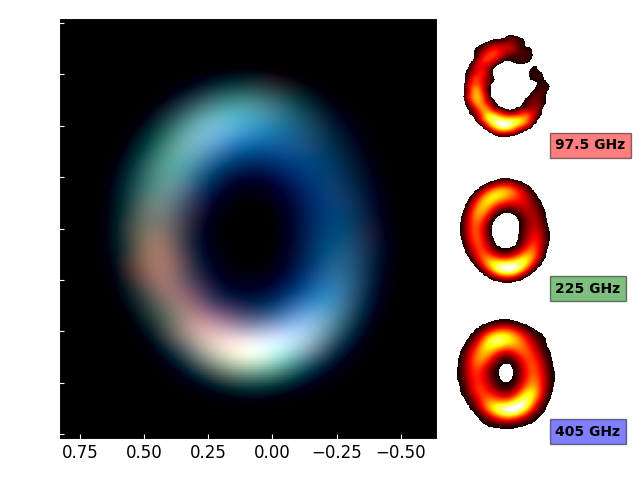}
  \caption{RGB image of the multi-frequency observations of
    ISO-Oph\,2\,A. Each image is masked below 10$\sigma$.  {\bf Red}:
    B3 continuum, restored with $r=0.7$, with a beam of
    $0\farcs230 \times0\farcs171$ along 90\,deg. {\bf Green}: B6
    continuum, degraded to the B3 beam. {\bf Blue}: B8 continuum,
    degraded to the B3 beam.  } \label{fig:rgb}
\end{figure}

\begin{figure}
  \centering
  \includegraphics[width=\columnwidth,height=!]{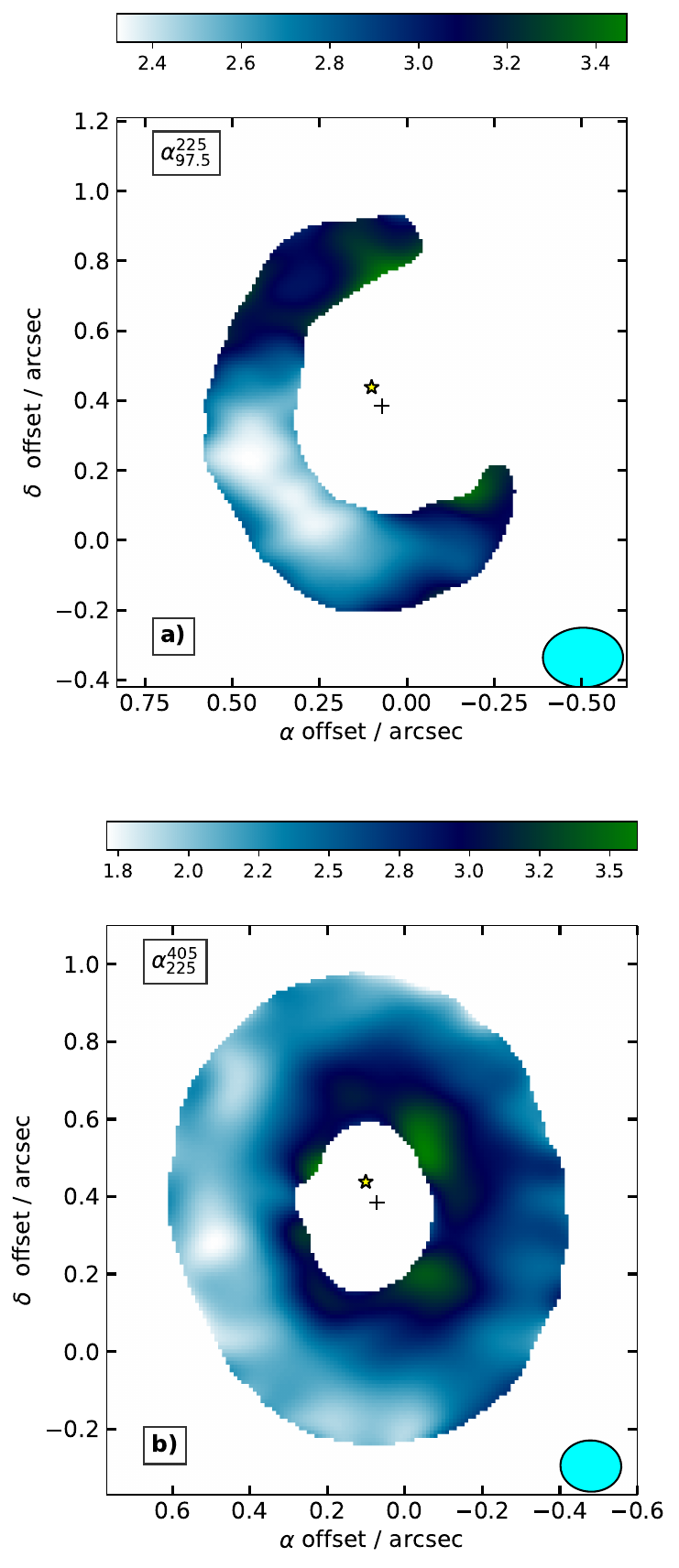}
  \caption{Intensity spectral index maps. {\bf a)} Spectral index between B3 and  B6, at the resolution of B3 ($r=0.7$).   {\bf b)} Spectral index between B6 and  B8, at the resolution of B8 ($r=0$).   } \label{fig:specindex}
\end{figure}

A summary of the self-calibrated and aligned data can be found in
Fig.\,\ref{fig:summary_mfreq}. The same figure but in brightness
temperature is provided in Appendix\,\ref{sec:Tbmaps}. A resolved
multi-frequency analysis requires all datasets to have a common
angular resolution. In Fig.\,\ref{fig:rgb} we compare B3 with degraded
version of B6 and B8, both smoothed to match B3. We also report in
Fig.\,\ref{fig:specindex} the intensity spectral index maps, in the
form
$\alpha_{\nu_1}^{\nu_2} =
\ln\left(I_{\nu_2}/I_{\nu_1}\right)/\ln\left(\nu_2/\nu_1\right)$.  The
structure of $\alpha_{97.5}^{225}$ and $\alpha_{225}^{405}$ are
remarkably different, which will be interpreted in
Sec.\,\ref{sec:analysis}.

%(base) simon@uaren:~/common/ppdisks/ISO-Oph_2/proc$ python match_zoom.py
%(base) simon@uaren:~/common/ppdisks/ISO-Oph_2/genfigs$ python summary_cont_mfreq.py
%(base) simon@uaren:~/common/ppdisks/ISO-Oph_2/report_mfreq$ rsync -va ~/common/ppdisks/ISO-Oph_2/genfigs/fig_mfreq_x3.pdf ./figs/
%(base) simon@uaren:~/common/ppdisks/ISO-Oph_2/report_mfreq$ rsync -va ~/common/ppdisks/ISO-Oph_2/genfigs/fig_mfreq_x4.pdf ./figs/
\begin{figure*}
  \centering
  \includegraphics[width=0.8\textwidth,height=!]{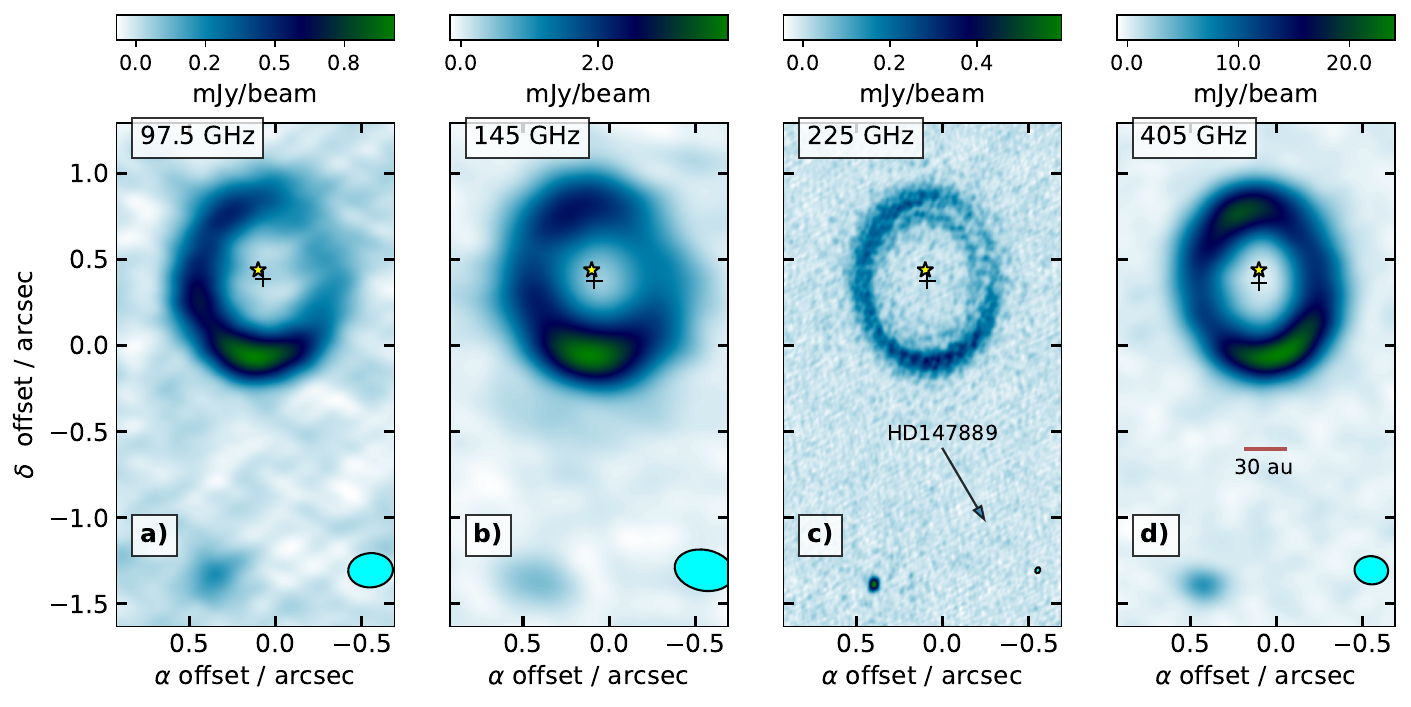}
  \caption{Summary of the multi-frequency observations of ISO-Oph\,2.
    The yellow star marks the position of ISO-Oph\,2A, while the plus
    sign marks the position of the ring centroid (see
    Sec.\,\ref{sec:orient}. ISO-Oph\,2B is the point source to the
    South. {\bf a}) B3 continuum, restored with $r=1$. {\bf b}) B4
    continuum, restored with $r=1$. {\bf c}) B6 continuum, restored
    with $r=2$. The arrow points to the direction of HD\,147889. {\bf
      d}) B8 continuum, restored with $r=1$. We provide a linear scale
    in au, assuming a distance of $134.3\pm7.7\,pc$
    \citep[][]{Gaia2022}.} \label{fig:summary_mfreq}
\end{figure*}

%(base) simon@uaren:~/common/ppdisks/ISO-Oph_2/genfigs$ python RGB_cont_mfreq.py
%(base) simon@uaren:~/common/ppdisks/ISO-Oph_2/report_mfreq$ rsync -va ~/common/ppdisks/ISO-Oph_2/genfigs/fig_RGB.png  ./figs/

\subsection{Astrometry of ISO-Oph\,2B}

The accuracy of the multi-frequency alignment is $\lesssim 9$\,mas,
which is slightly better than the rule-of-thumb for the ALMA pointing
accuracy, of about $1/10$ of the clean beam \citep[][]{ALMA_technical_handbook2019athb.rept.....R}. In
Table\,\ref{table:orbitB} we record the positions of ISO-Oph\,2B,
measured with elliptical Gaussian fits. The error budget is dominated
by that of the Gaussian centroid, except for B6, for which we assign
the nominal ALMA pointing accuracy.

It appears that ISO-Oph\,2B is moving too fast, relative to
ISO-Oph\,2A, for Keplerian rotation. At their projected separation, of
$\sim 257\,$au, the Keplerian velocity for a $0.5\,M_\odot$ system is
$\sim$1.3\,km\,s$^{-1}$, and only $\sim$0.8\,km\,s$^{-1}$ after
projection onto the plane of the sky with a disk inclination of
36\,deg. However, the projected velocity comparing SB17 and B8 is
$10.1\pm3.4$\,km\,s$^{-1}$, and is $6.0\pm2.5$\,km\,s$^{-1}$ when
comparing B6 and B8. The difference with a bound orbit and circular
orbit in the plane of the circum-primary disk are $\sim2.7\sigma$ and
$\sim$2.1$\sigma$. A new epoch is required to conclude. 

To further assess the possibility of an unbound trajectory for ISO-Oph 2B, we attempted to fit the 5 astrometric measurements with a bound orbit using {\sc Orbitize!} \citep{Blunt2020}. We assumed a total mass of $0.58\pm0.15 M_{\odot}$ for the system \citep[0.5 $M_\odot$ for A and 80 $M_J$ for B;][]{Gonzalez-Ruilova2020ApJ...902L..33G}, and a parallax of $7.449\pm0.074$ mas \citep%[from {\em GAIA}
%DR\,3,][]
{Gaia2016, Gaia2022}.
We considered two cases: no prior on the orbit, and tight Gaussian priors on the inclination and longitude of the ascending node for the orbital plane to match the plane of the circumprimary disc. In either cases, %a relatively uniform posterior was found for eccentricity after 10,000 orbits were drawn with the OFTI algorithm. We also 
we drew 10,000 orbits with the OFTI algorithm and 
noted 
%in either cases 
that the first epoch datum was $\sim$2 $\sigma$ diskrepant from the closest orbit predictions at that epoch out of these 2 $\times$ 10,000 samples. This provides another piece of evidence in favour of an unbound hyperbolic trajectory (i.e.~a fly-by).%, although a new astrometric point is desirable to unambiguously confirm this hypothesis.

%***PENDING: these velocity differences are close to 3sigma. Perhaps an orbital fit could convincingly confirm an hyperbolic orbit? **** 

% ~/common/ppdisks/ISO-Oph_2/proc/README_B_orbit

\begin{table}
  \caption{Multi-epoch astrometry for ISO-Oph\,2B, relative to the phase-center for B6. In this system, the {\em GAIA} coordinates for the primary are
    $\Delta \alpha  = 0\farcs101$, $\Delta \delta = 0\farcs438$.  }
 \label{table:orbitB}
 \begin{tabular}{cccc}
  \hline
   Date & Dataset   & $\Delta \alpha^a$  & $\Delta \delta^b$ \\ \hline
   2017-07-13  & SB17 & $0\farcs363\pm0\farcs023$ &$-1\farcs349\pm0\farcs023$ \\
   2019-06-12  & B6 & $0\farcs399\pm0\farcs003$ &$-1\farcs385\pm0\farcs003$ \\
   2021-07-27  & B3 & $0\farcs367\pm0\farcs028$ &$-1\farcs341\pm0\farcs028$ \\
   2022-07-20  & B4 & $0\farcs423\pm0\farcs024$ &$-1\farcs368\pm0\farcs024$ \\   
   2022-08-04  & B8 & $0\farcs426\pm0\farcs013$ &$-1\farcs393\pm0\farcs013$ \\  
 \end{tabular}\\
 $^a$~Offset along R.A., in arcsec.
 $^b$~Offset along Dec., in arcsec.
\end{table}

\subsection{Photometry}

Table\,\ref{table:phot} reports the integrated flux densities for each component 
of ISO-Oph\,2. For ISO-Oph\,2A we used aperture photometry within a radius of $0\farcs8$, centered on the primary. For ISO-Oph\,2B we used the integrated flux density obtained with elliptical Gaussian fits. 

% generated with python summary_cont_mfreq_wB3.py
\begin{table}
  \caption{Photometry of ISO-Oph\,2 for each of the datasets presented in Fig.\,\ref{fig:summary_mfreq}. We report flux densities in mJy. The errors do not include the systematic calibration uncertainty.  }
 \label{table:phot}
 \begin{tabular}{ccccc}
  \hline
 & B3 & B4 & B6 & B8  \\ \hline
ISO-Oph\,2B  & $0.32\pm0.03$ & $0.50\pm0.07$ & $1.81\pm0.01$  &  $7.3\pm0.3$ \\
ISO-Oph\,2A  & $6.05\pm0.15$ & $23.6\pm0.3$ & $85.5\pm0.5$ &  $316\pm2$ \\
 \end{tabular}
\end{table}

\subsection{Disk orientation and stellar offset} \label{sec:orient}

The position of ISO-Oph\,2A \citep[from {\em GAIA}
DR\,3,][]{2020yCat.1350....0G}, corrected for proper motion, is at
$\Delta \alpha = 0\farcs101$ and $\Delta \delta = 0\farcs438$ relative
to the B6-LB19 phase center. The errors on these coordinates are
negligible relative to the ALMA pointing accuracy for B6-LB19, which
is $\sim$3\,mas. ISO-Oph\,2A is remarkably offset from the ring
centroid. Fig.\,\ref{fig:summary_mfreq} indicates the stellar position
and the cavity centers for B6 and B8. We first centered each image on
the primary, and then estimated the disk orientation and center using
the {\sc MPolarMaps} package \citep[described in
][]{Casassus2021MNRAS.507.3789C}. {\sc MPolarMaps} minimizes the
azimuthal scatter in radial profiles, which should yield the correct
orientation parameters for an axially symmetric disk. Under this
assumption, the best fit parameters for B6 are a position angle
PA$ = 2.42^{+0.45}_{-0.49}$\,deg, an inclination of
$i = 36.92^{+0.21}_{-0.24}$\,deg, and disk center relative to the
stellar position: $\Delta \alpha = -0\farcs011\pm0\farcs001$
$\Delta \delta = -0\farcs062\pm0\farcs001$. The corresponding PA and
$i$ are consistent with those reported by
\citet{Gonzalez-Ruilova2020ApJ...902L..33G}. For B4, we obtain  PA$ = 12.4^{+1.4}_{-1.4}$\,deg,
$i = 31.8 ^{+0.8}_{-0.9}$\,deg, and
$\Delta \alpha = -0\farcs015\pm0\farcs002$
$\Delta \delta = -0\farcs065\pm0\farcs002$, while for B8,
PA$ = 7.8^{+0.58}_{-0.65}$\,deg, an inclination of
$i = 36.62 ^{+0.41}_{-0.38}$\,deg, and disk center relative to the
stellar position: $\Delta \alpha = -0\farcs001\pm0\farcs001$
$\Delta \delta = -0\farcs076\pm0\farcs001$.

\begin{figure*}
  \centering
  \includegraphics[width=0.6\textwidth,height=!]{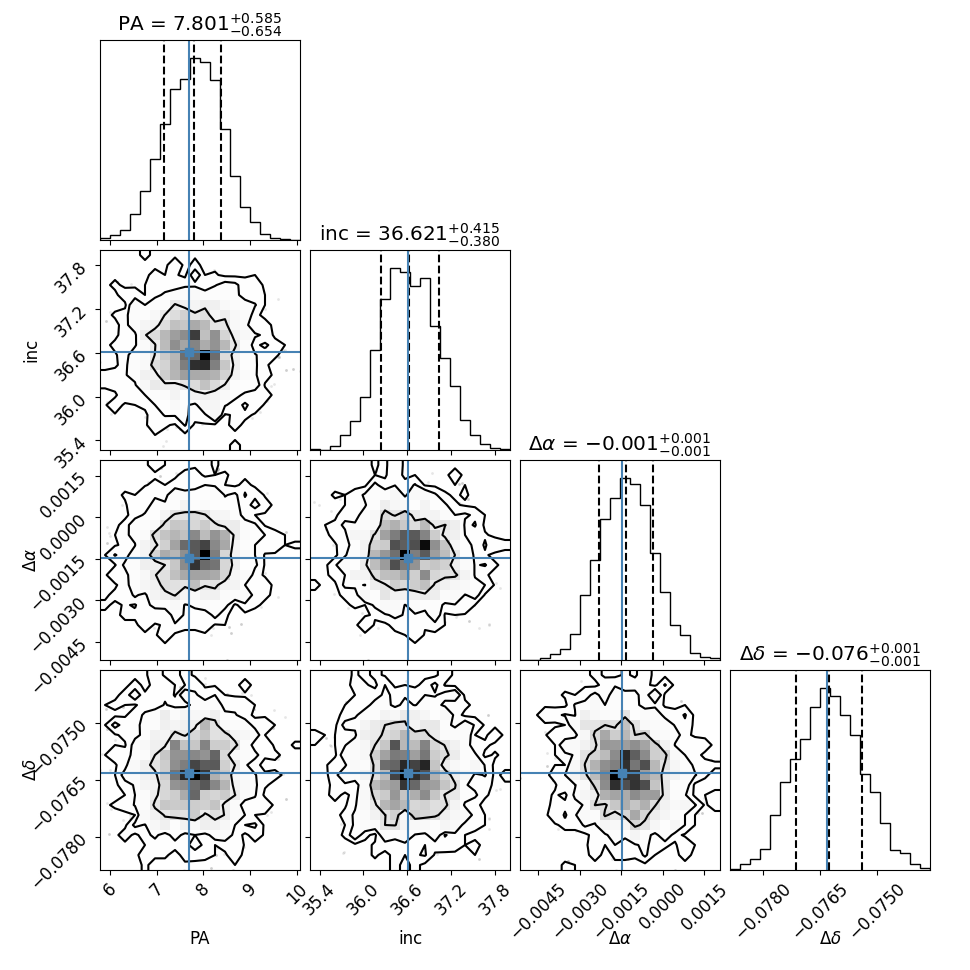}
  \caption{Corner plot from the optimisation of the
  orientation parameters in ISO-Oph\,2\,A, using the B8 image with $r=0$. The histograms plot the 1D
  probability density functions of the parameters indicated in titles,
  along with their median values and 1$\sigma$ confidence intervals
  (i.e. at 16\% and 84\%), which are also shown by the vertical dashed
  lines. The contour plots show the marginalized 2D distributions
  (i.e. the 2D projection of the 8D posterior probability
  distribution), for the corresponding pairs of parameters. Contour
  levels are chosen at 0.68,  0.95 and 0.997.
  } \label{fig:polarmapsb8}
\end{figure*}

In terms of their posterior distributions, the disk orientations
inferred from {\sc MpolarMaps} are well constrained (see
Fig.\,\ref{fig:polarmapsb8}). However, the above errors do not
consider the systematics induced by the non-axial symmetry of the
disk, which could well be intrinsically eccentric, hence the
significant differences for each image. Still, a qualitatively large
stellar offset, which can readily be seen by eye, is common to all 3
images. Such a large offset, of $\sim 62$ to $\sim76$\,mas, is rare in
ringed systems with accurate optical/IR stellar astrometry. The
associated eccentricity, for a 0\farcs43 ring, ranges between 0.12 and
0.17.  By comparison, the largest of such offsets, inferred from
long-baseline continuum ALMA data, is $33 \pm 3$\,mas in MWC\,758
\citep[][]{Dong2018ApJ...860..124D}, and $12 \pm 4$\,mas in
HD\,135344B \citep[][]{Casassus2021MNRAS.507.3789C}. The ring around
IRS\,48 also appears to be extremely eccentric, with $e\sim0.27$ and
an offset between the ring centroid and the central sub-mm emission of
$\sim$0\farcs15 \citep{Yang2023ApJ...948L...2Y}.

\section{Analysis} \label{sec:analysis}

\subsection{Qualitative spectral trends}

The RGB image in Fig.\,\ref{fig:rgb} is rich in structure, suggesting
strong azimuthal variations in physical conditions. The low frequency
spectral index $\alpha_{97.5}^{225}$ in Fig.\,\ref{fig:specindex}a
reveals a clump of values around $\sim 2.4$ to the South-East, which
could be the result of optically thin emission from a concentration of
large grains, while the rest of the disk corresponds to smaller grains
with $\alpha_{97.5}^{225} \sim 3.0$. In turn $\alpha_{225}^{405} $ in
Fig.\,\ref{fig:specindex}b is more uniform with values of around 2.0
to the East, which could correspond to optically thick emission, while
emission on the Western side is more optically thin, with
$\alpha_{225}^{405} \sim 3$. An optically thick region to the East
could correspond to a lopsided disk, which would be consistent with
the large grain clump in the dust trap hypothesis: grains with larger
dimensionless stopping time (Stokes number), up to $S_t \lesssim 1$,
pile up near the center of the pressure maximum
\citep[e.g.][]{Birnstiel2013A&A...550L...8B,
  LyraLin2013ApJ...775...17L, Zhu_Stone_2014ApJ...795...53Z,
  Mittal2015ApJ...798L..25M, BZ2016MNRAS.458.3927B,
  Casassus2019MNRAS.483.3278C}.

\subsection{Implementation of  uniform-slab diagnostics in {\sc Slab.Continuum}}

We quantify the dust trapping scenario by interpreting the spectral
variations in terms of dust properties averaged along the line of
sight.  We developed the package {\sc Slab.Continuum} to model the
emergent intensities from a uniform-slab, including isotropic
scattering \citep[in the Eddington approximation and with two-streams
boundary conditions, following
][]{MiyakeNakagawa1993Icar..106...20M,dAlessio2001ApJ...553..321D,Sierra2017ApJ...850..115S,Sierra2019ApJ...876....7S,Casassus2019MNRAS.483.3278C}:
\begin{equation}
  I^m_\nu(\tau_\nu,\mu)  =   B_\nu(T)\left[  1- \exp\left(-\frac{\tau_\nu}{\mu}\right) + \omega_\nu \mathcal{F_\nu} \right] \label{eq:Iemerge}
\end{equation}
where
\begin{equation}
\begin{split}
  \mathcal{F_\nu}  = & \frac{1}{  (\sqrt{1 - \omega_\nu} - 1) \exp\left(-\sqrt{ 3 (1-\omega_\nu)}\tau_\nu \right) - (\sqrt{1-\omega_\nu} + 1) } \times  \\
 &  \left\{   \frac{ 1- \exp\left[ - \left(  \sqrt{3(1-\omega_\nu)} + \mu^{-1}  \right) \tau_\nu \right] }{ \sqrt{3(1-\omega_\nu)} \mu + 1} ~ + \right. \\
& \left. \frac{ \exp\left( -\frac{\tau_\nu}{\mu}  \right) - \exp\left( -\sqrt{3(1-\omega_\nu)} \tau_\nu \right)    }{ \sqrt{3(1-\omega_\nu)} \mu - 1}  \right\} , \label{eq:slab}
\end{split}
\end{equation}
$\omega = \frac{\kappa^{\rm sca}}{\kappa^{\rm abs} +\kappa^{\rm sca}}$
is the dust albedo, $\tau_\nu \equiv \Sigma_g \kappa_\nu$, and
$\kappa_\nu = \kappa^{\rm abs} +\kappa^{\rm sca}$. The angle of
incidence, $\mu = \cos(i)$, was set to 1 for simplicity, thus
reducing Eq.\,\ref{eq:slab} to Eqs. 24 and 25 of  \citet{Sierra2019ApJ...876....7S}.

The size-averaged dust opacities $\kappa_\nu^{\rm abs}$ and
$\kappa_\nu^{\rm sca}$ were computed using routines from the {\sc
  dsharp\_opac}
package\footnote{\url{https://github.com/birnstiel/dsharp_opac}},
described in \citet[][]{Birnstiel2018ApJ...869L..45B}, and with their
default effective optical constants \citep[Fig.\,2 in
][, i.e. ``DSHARP'' opacities]{Birnstiel2018ApJ...869L..45B}.  Forward scattering was accounted
for by correcting the scattering opacity $\kappa_\nu^{\rm sca}$ to
$(1 - g_\nu) \kappa_\nu^{\rm sca}$, where $g_\nu$ is the Henyey-Greenstein
anisotropy parameter.

For a power-law distribution of dust grain sizes, with a single dust
composition, and for a fixed gas-to-dust mass ratio (taken here to be
100), the free-parameters for any given line-of-sight are the total
mass column density $\Sigma_g$, the maximum grain size $a_{\rm max}$,
the dust size exponent $q$, and the dust temperature $T_{\rm d}$. We fit the
spectral-energy-distribution (SED) for each line-of-sight,
with $N$ independent frequency points $\{I_{\nu_i} \}_{i=1}^N$,  by minimizing 
\begin{equation}
  \chi^2 = \sum_i \frac{(I_{\nu_i}- I^m_{\nu_i})^2}{\sigma_i^2}, \label{eq:chi2}
\end{equation}
Where the weights $\{1/\sigma_i^2\}$ are approximated as the
root-mean-square dispersion for each residual image\footnote{the
  dirty map of the residual visibilities}, including the flux
calibration error in quadrature. The flux calibration accuracy was
taken to be 5\% in B3, 5\% in B4,  5\% in B6, and 10\% in B8.

The posterior disributions were calculated with a Markov chain Monte
Carlo ensemble sampler \citep{MCMC2010CAMCS...5...65G}. We used the
{\tt emcee} package \citep[][]{emcee2013PASP..125..306F}, with 1000
iterations, a burn-in of 800, and 10 walkers per
free-parameter. Except for $q$, we varied the logarithm of each
parameter, with flat priors, and across wide domains in parameter
space: $ 0 < \log(T_{\rm d}/{\rm K}) <3 $,
$-5 < \log(\left(\Sigma_g / {\rm g\,cm}^{-2}\right) < 3$ and
$-3 < \log\left(a_{\rm max}/{\rm cm}\right)< 10$, and
$-3.99 < q < -2$. The {\sc Slab.Continuum} package optionally runs a
final optimization with the Powell variant of the conjugate-gradient
minimization algorithm, using the maximum likelihood parameters
obtained with {\tt emcee}. Rather than calculate size-averaged
opacities for all sampled values of $a_{\rm max}$ and $q$, we first
computed opacity grids in $a_{\rm max}$ and $q$, at each of the
frequencies $\{ \nu_i \}_{i=1}^N$, and used bi-linear interpolation.

\subsection{Application of {\sc Slab.Continuum} to ISO-Oph\,2A}

Initial trials including B4 resulted in strong biases due to beam
dilution, when the beam is much larger than the structures (see
below). We therefore discarded B4 from the spectral fits. With $N=3$
independent spectral points we may optimize only up to 3
parameters. For the present application of {\sc Slab.Continuum} we
thus chose to fix $q=-3.5$, i.e. as in the standard ISM size
distribution \citep[][]{Mathis1977ApJ...217..425M}.

Before running the optimization on all lines-of-sights, and in order
to reduce the load on computer resources, we resampled each image
$\{I_{\nu_i} \}_{i=1}^N$ into coarser pixels (in synthesis imaging the
pixel size is usually chosen to be around 1/10 of the natural-weights
clean beam). We additionally set an intensity mask at $10\sigma$ in
B6.

\begin{figure*}
  \centering
  \includegraphics[width=0.6\textwidth,height=!]{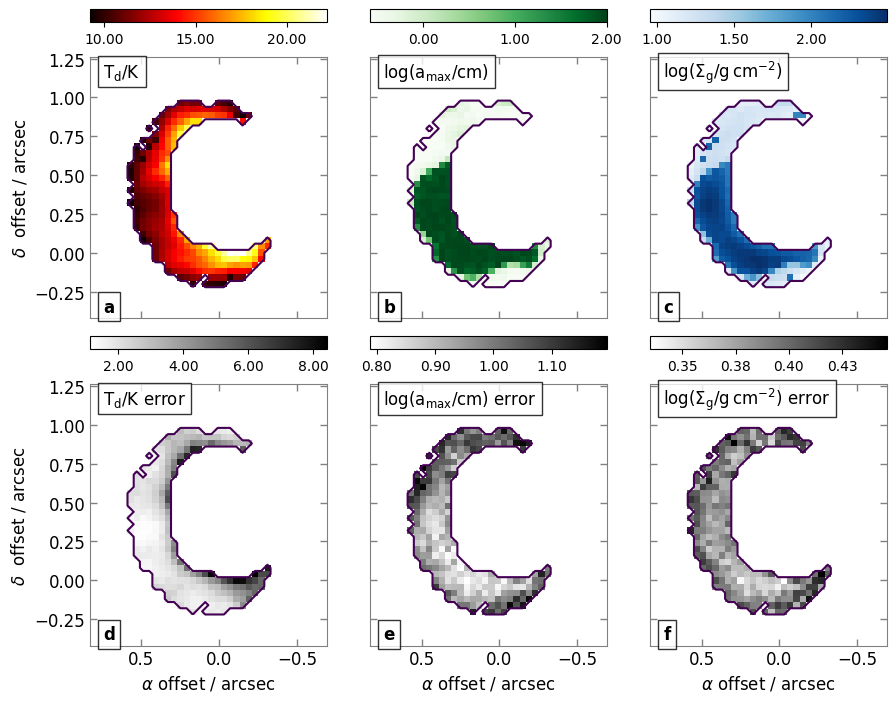}
  \caption{Dust parameters for each line of sight from the
    uniform-slab optimizations of $T_{\rm d}$, $a_{\rm max}$ and
    $\Sigma_g$, and constrained with the B3, B6 and B8 all degraded to
    the B3 beam with $r=0.7$. The black contours outline an error mask
    on $\log_{10}(T_{\rm d})$, set to 0.2. {\bf a}: Maximum-likelihood
    $T_{\rm d}$. {\bf b}: Maximum-likelihood $\log_{10}(a_{\rm
      max})$. {\bf c}: Maximum-likelihood $\log_{10}(\Sigma_g)$.  {\bf
      d}: One-sigma uncertainty on $T_{\rm d}$, approximated as
    $\sigma(T_{\rm d}) = \ln(10) T_{\rm d} \sigma(\log(T_{\rm d}))$,
    where $\sigma(\log(T_{\rm d}))$ is estimated with the average of
    the upwards and downwards 34\% confidence intervals around the
    median. {\bf e}: One-sigma uncertainty on
    $\log_{10}(a_{\rm max})$. {\bf f}: One-sigma uncertainty on
    $\log_{10}(\Sigma_g)$. The blue lines correspond to the
    maximum-likelihood values.  } \label{fig:imoptim_b3_b6_b8}
\end{figure*}

\begin{figure}
  \centering
  \includegraphics[width=\columnwidth,height=!]{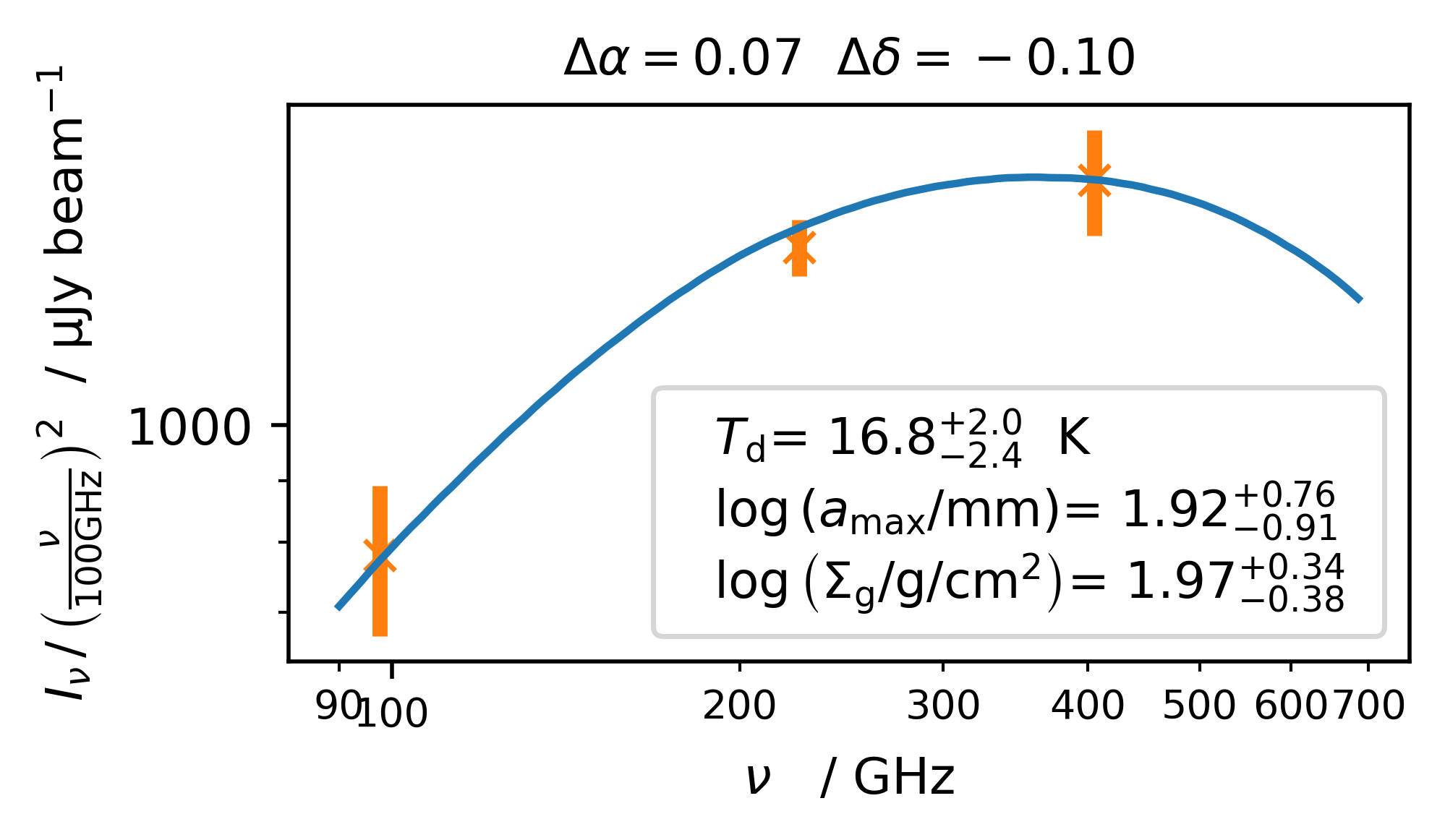}
  \includegraphics[width=\columnwidth,height=!]{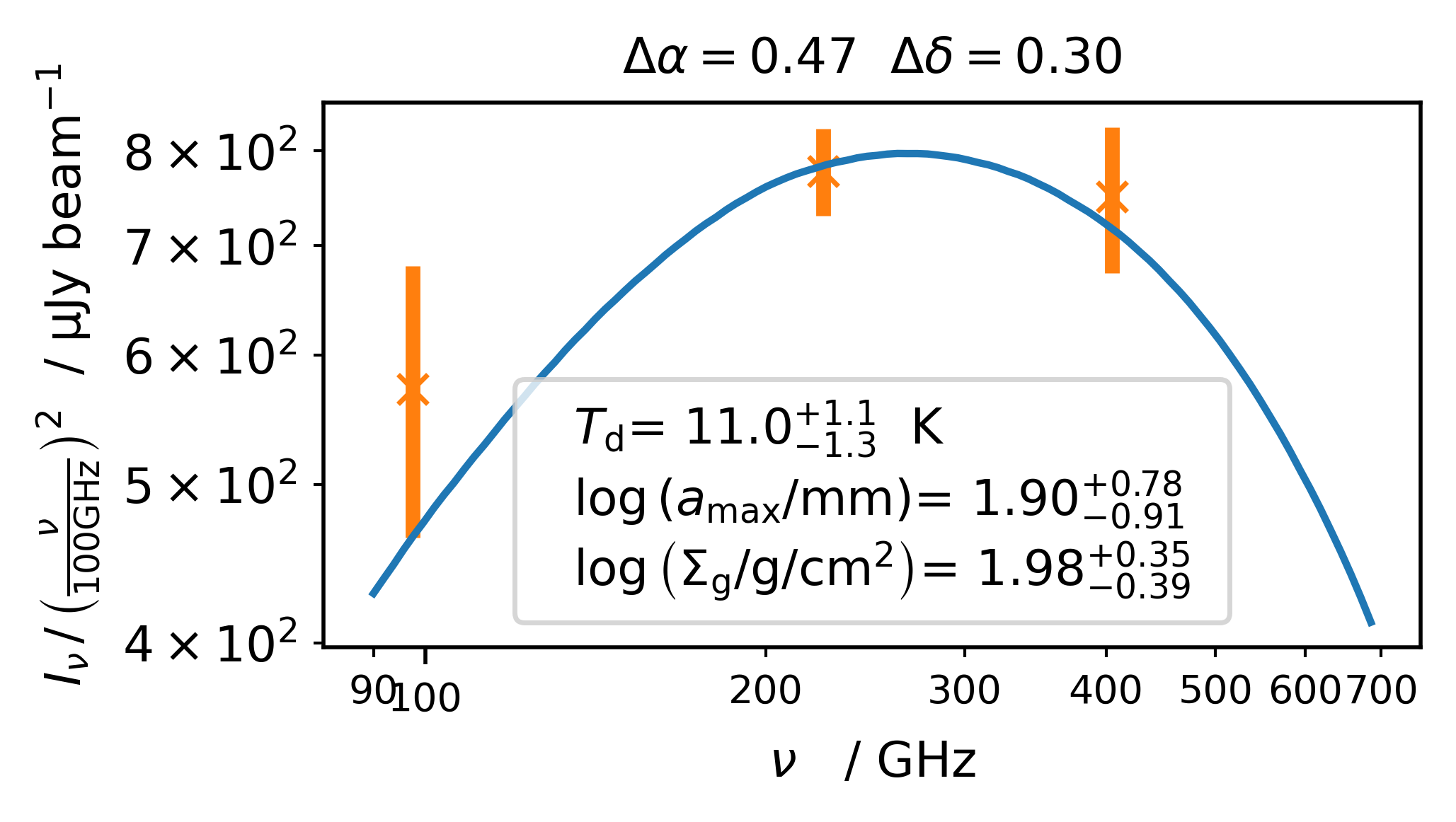}
  \caption{Spectral energy distribution and best-fit uniform-slab model for two example lines of sight (LOS). The direction of each LOS is given as  offset from the phase centers, in arcsec,  on top of each plot.     {\rm Top}: LOS  towards the peak B6 intensity. {\rm Bottom}: LOS  towards the minimum in  B6 intensity along the eastern side of the ring.} \label{fig:SEDs}
\end{figure}

\begin{figure*}
  \centering
  \includegraphics[width=0.49\textwidth,height=!]{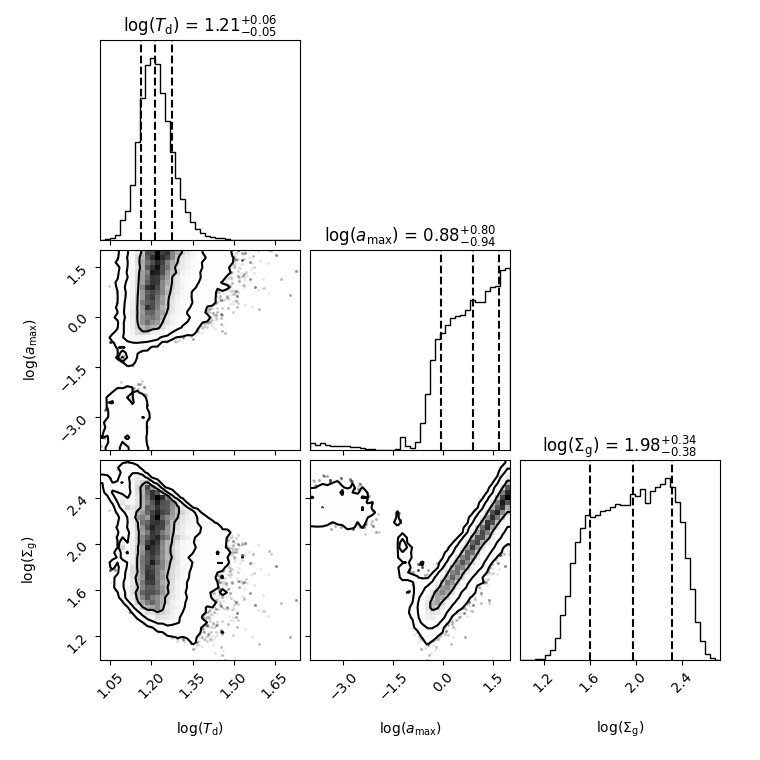}
  \includegraphics[width=0.49\textwidth,height=!]{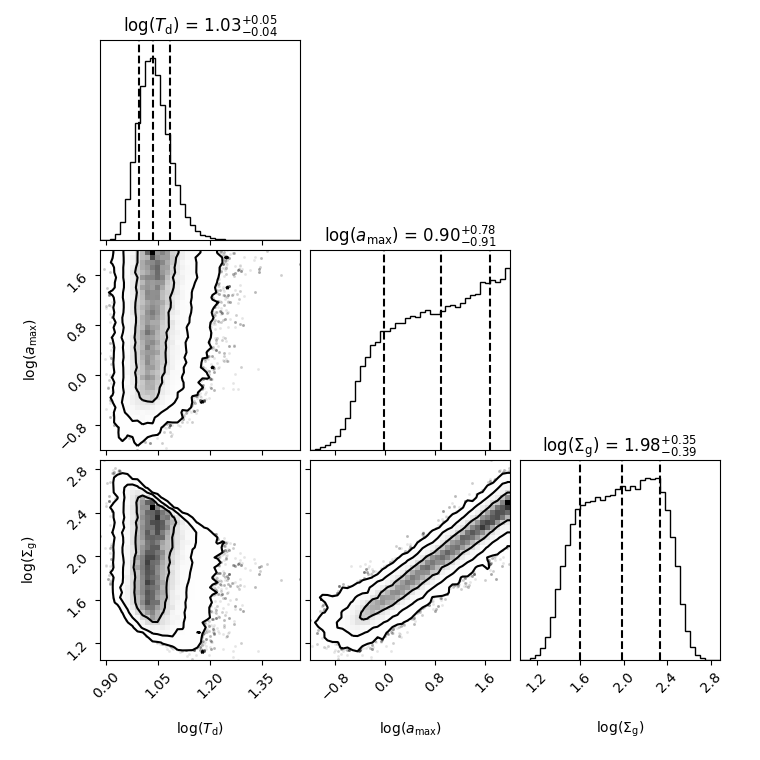}
  \caption{Annotations follow from Fig.\,\ref{fig:polarmapsb8}. {\bf
      Left}: Corner plot for the dust parameters towards the peak B6
    intensity, corresponding to Fig.\,\ref{fig:SEDs} (top). {\bf
      Left}: Corner plot for the dust parameters towards the minimum
    B6 intensity along the Eastern ring, and corresponding to
    Fig.\,\ref{fig:SEDs} (bottom).  } \label{fig:corner}
\end{figure*}

The coarsest data-point is B3, and thus the full set of frequency
points was degraded to the B3 beam with $r=0$. The result is
summarised in Fig.\,\ref{fig:imoptim_b3_b6_b8}, where we see that the
inferred dust parameters bear fairly large uncertainties, especially
to the West, where the disk is faintest in B3. Example SEDs, for the
lines-of-sight towards the intensity extrema in B6 and along the
eastern side of the ring, are shown in Fig.\,\ref{fig:SEDs}. A
corner-plot for the posterior probability distributions towards the
peak is shown in Fig.\,\ref{fig:corner}.  We see that $a_{\rm max}$ is
poorly constrained, which we tentatively interpret in terms of two
dominant effects. First, towards the minimum in B4, to the West of the
ring, the solutions for large grains,
$\log(a_{\rm max}/{\rm cm}) > 0.8$, correspond to optically thin
emission at all frequencies, where intensities are proportional to
optical depth, $I_\nu \propto T_{\rm d} \tau_\nu$. Since
$\tau_\nu \propto \Sigma_g$, the lack of an additional point in the
optically thick regime prevents lifting the $T_{\rm d} - \Sigma_g$
degeneracy. Second, even with an optically thick point with which to
set $T_{\rm d}$, for grains that are larger than the wavelength,
increasing grains larger have lower opacity $\kappa_\nu$, so that a
given optical depth may result from arbitrarily large grains
$\tau_\nu = \kappa_\nu \, \Sigma_g$, leading to the
$a_{\rm max} - \Sigma_g$ degeneracy
\citep[e.g.][]{Sierra2021ApJS..257...14S}. We
set the maximum grain size to 100\,cm, which corresponds to the
maximum size in the default opacities from
\citet[][]{Birnstiel2018ApJ...869L..45B}.

%Indeed, the well known
%two-size degeneracy, by which there are two solutions for
%$a_{\rm max}$ for a wide range of spectral indices (at a single
%frequency), maps the bimodal probability density function (PDF) in
%$a_{\rm max}$ into a corresponding bimodal PDF in $\Sigma_g$ (see
%Fig.\,\ref{fig:corner}).

% (base) simon@uaren:~/common/ppdisks/ISO-Oph_2/mfreqfits$ python ./gen_summary_dustparams_wb3r07.py
% rsync -va ~/common/ppdisks/ISO-Oph_2/mfreqfits/output_imoptim_wb3r07/fig_dustparams_linear.png ./figs/fig_dustparams_b3r07_linear.png
% rsync -va ~/common/ppdisks/ISO-Oph_2/mfreqfits/output_imoptim_wb3r07/fig_dustparams.png ./figs/fig_dustparams_b3r07.png

%(base) simon@uaren:~/common/ppdisks/ISO-Oph_2/report_mfreq$ rsync -va ~/common/ppdisks/ISO-Oph_2/mfreqfits/output_imoptim_wb4r0/fig_bestfit.png ./figs/
%(base) simon@uaren:~/common/ppdisks/ISO-Oph_2/report_mfreq$ rsync -va ~/common/ppdisks/ISO-Oph_2/mfreqfits/output_imoptim_wb4r0/triangle_all.png  ./figs/

% rsync -va   ~/common/ppdisks/ISO-Oph_2/mfreqfits/output_imoptim_wb3r07/fig_bestfit_minB6.png  ./figs/fig_bestfit_minB6.png
% rsync -va   ~/common/ppdisks/ISO-Oph_2/mfreqfits/output_imoptim_wb3r07/triangle_all.png  ./figs/triangle_all_minB6.png
% rsync -va   ~/common/ppdisks/ISO-Oph_2/mfreqfits/output_imoptim_wb3r07/fig_bestfit_peakB6.png  ./figs/fig_bestfit_peakB6.png
% rsync -va   ~/common/ppdisks/ISO-Oph_2/mfreqfits/output_imoptim_wb3r07/triangle_all.png  ./figs/triangle_all.png

%36\%, 50\%, and 84\%. The countours are drawn at 0.68, 0.95, 0.997.

\subsection{Incorporation of a beam filling-factor}

% 
%a strong correlation with $\Sigma_g$.

Beam-dilution, or the reduction of specific intensities in sources
that do not fill the clean beam, is strong in the coarse B3 beam, even
with $r=0$. This also translates into a reduction in brightness
temperature, which reach only $\sim$5\,K in B3, and $\sim$10\,K in the
finer beam of B6 (in ISO-Oph\,2A, Fig.\,\ref{sec:Tbmaps}).  Without a
beam filling factor among the free-parameters, the uniform-slab model
is a poor approximation. For instance, in the case of $T_{\rm d}$ and
optically thick emission, beam-dilution would lower the brightness
temperature, but the spectral indices would still correspond to the
undiluted black-body emission.  Similarly, $a_{\rm max}$ determines
the spectral indices and opacity, which may not match the observed
intensities if they are diluted. A question  arises on the
impact of beam-dilution on the inferred physical parameters.

\begin{figure}
  \centering
  \includegraphics[width=\columnwidth,height=!]{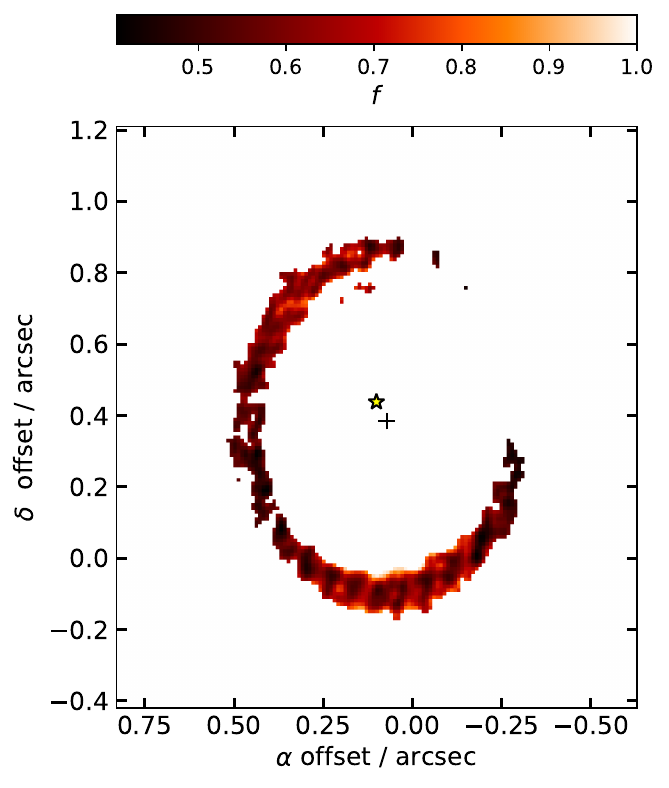}
  \caption{Beam filling factor inferred from the B6
    dataset.} \label{fig:fillfactor}
\end{figure}

\begin{figure*}
  \centering
  \includegraphics[width=0.6\textwidth,height=!]{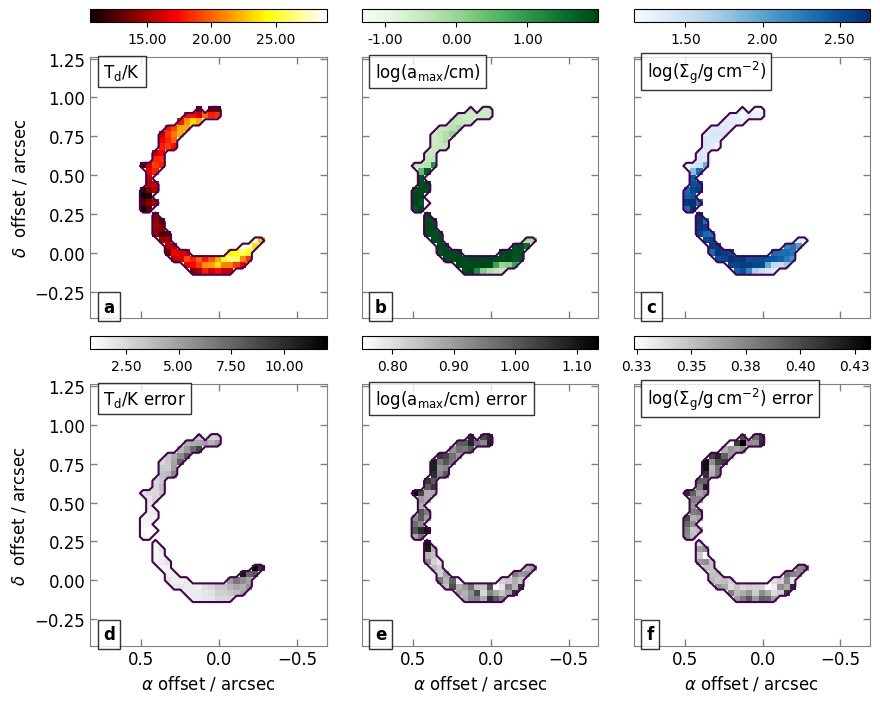}
  \caption{Dust parameters for each line of sight from the
    uniform-slab optimizations of $T_{\rm d}$, $a_{\rm max}$ and
    $\Sigma_g$, including a filling-factor, and constrained with the
    B3, B6 and B8 all degraded to the B3 beam with
    $r=0.7$. Annotations follow from
    Fig.\,\ref{fig:imoptim_b3_b6_b8}. } \label{fig:imoptim_b3_b6_b8_wfillfactor}
\end{figure*}

\begin{figure}
  \centering
  \includegraphics[width=\columnwidth,height=!]{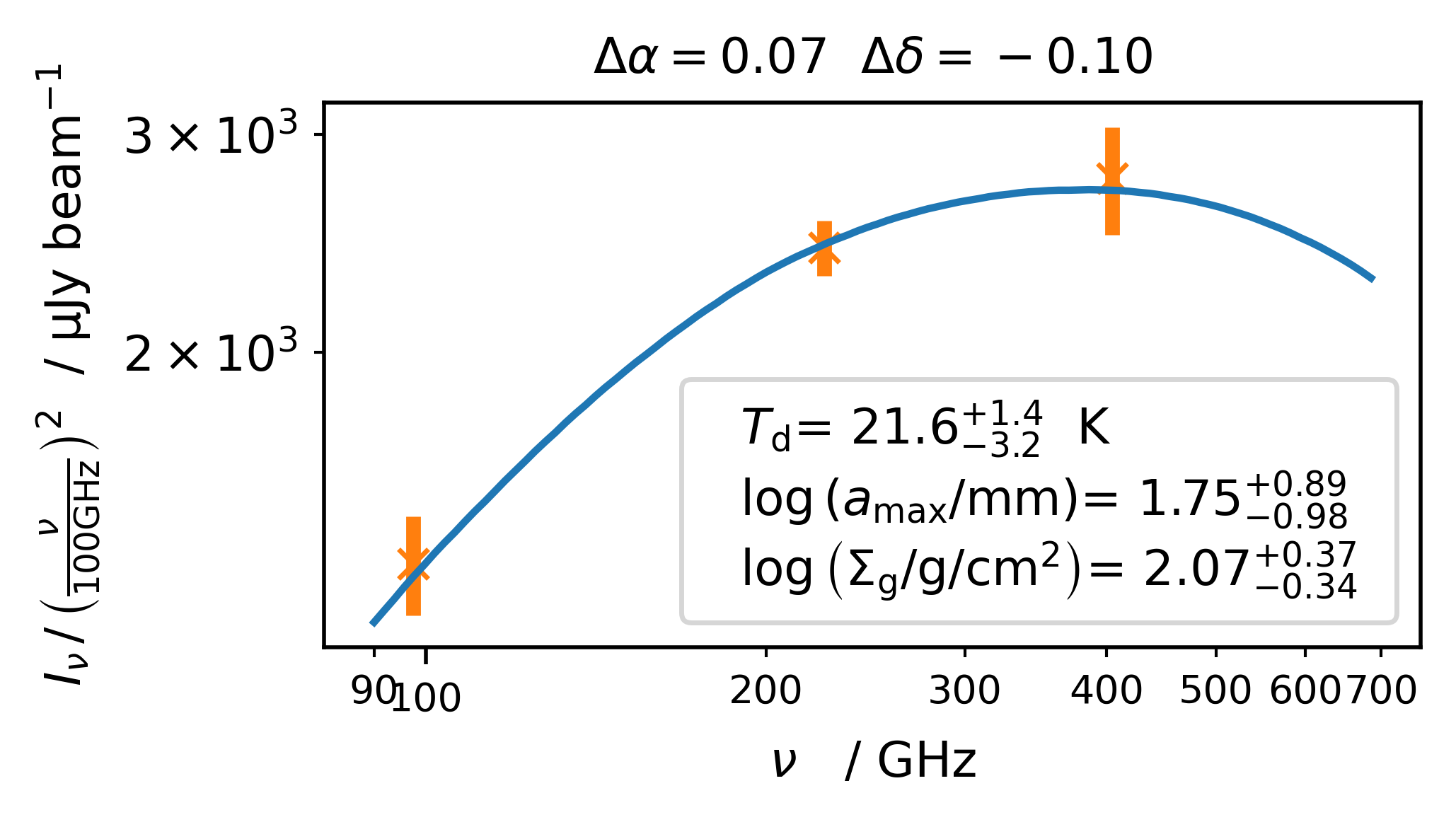}
  \includegraphics[width=\columnwidth,height=!]{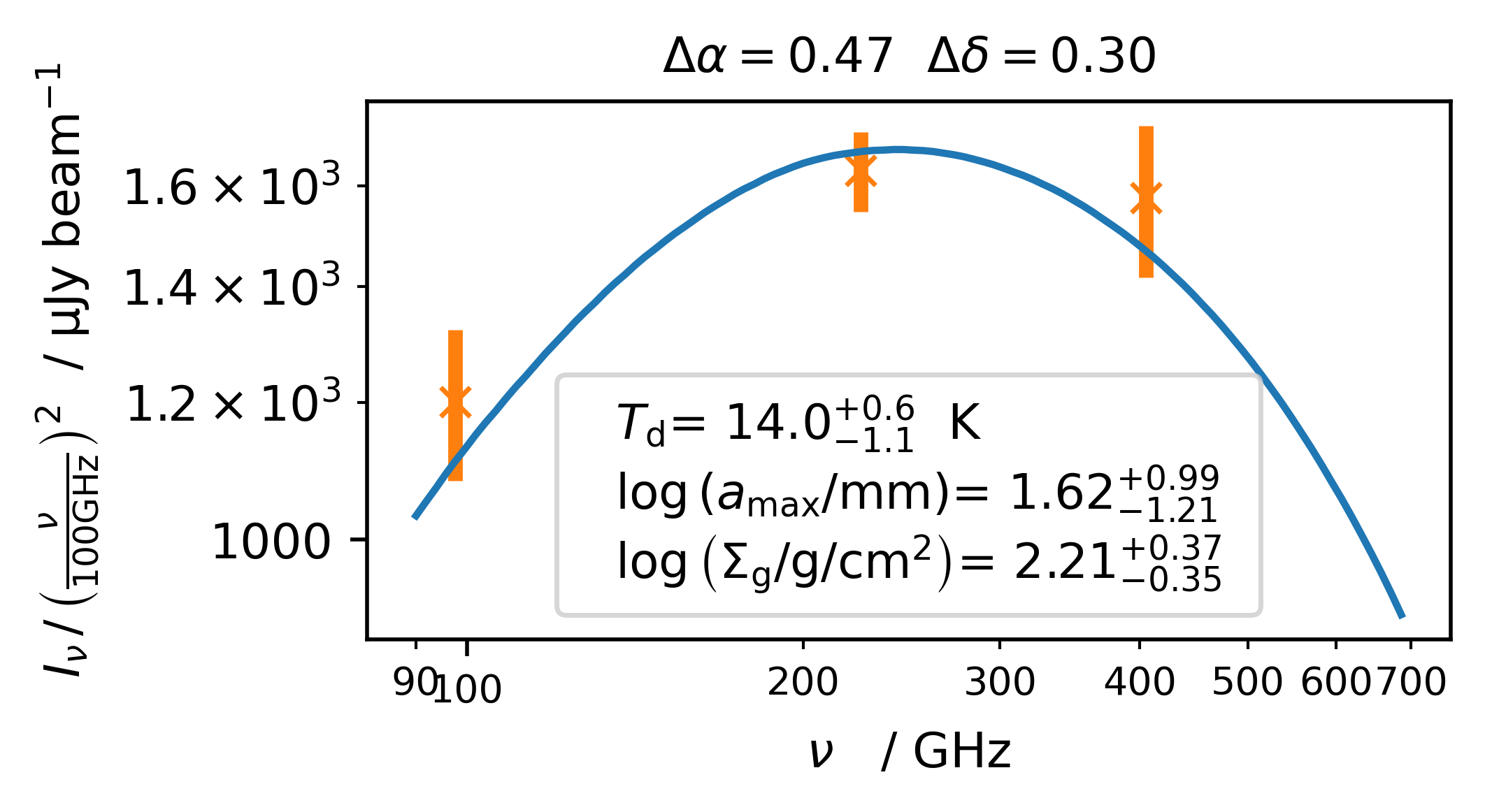}
  \caption{Spectral energy distribution and best-fit uniform-slab model for two example lines of sight (LOS), with a filling factor. Annotations follow from Fig.\,\ref{fig:SEDs}.  } \label{fig:SEDswfillfactor}
\end{figure}

The B6 dataset is well-sampled, and can be used to estimate the
filling factor in B6, by comparing the B6 map in native resolution
($I_{\rm B6}$) with its smoothed version ($I_{\rm B6}^{\rm s}$),
$f = I_{\rm B6}^{\rm s} / I_{\rm B6} $, with an upper limit of 1. The
resulting map is shown in Fig.\,\ref{fig:fillfactor}. We use this map
to scale the multi-frequency intensities, i.e. the corrected
intensities are $I_\nu^c = I_\nu / f$. The corresponding dust
properties are shown in Fig.\,\ref{fig:imoptim_b3_b6_b8_wfillfactor} ,
with examples SEDs in Fig.\,\ref{fig:SEDswfillfactor}. It is
interesting to note that $\chi^2$, as given by Eq.\,\ref{eq:chi2}, is
reduced from 1.1 to 0.11 with the incorporation of the filling factor
for the line of sight toward the peak B6 intensity. This probably
reflects the improved model in B6. However, there is no appreciable
difference for the line of sight towards the minimum B6 intensity (to
the East of the ring), where $\chi^2$ increases from 1.1 to 1.2 with
the inclusion of a filling factor.

%nofillfactor:
%minB6
%running final lnlike to set model to best L  values
%chi2 = 1.098219e+00
%peakB6:
%running final lnlike to set model to best L  values
%chi2 = 1.623038e-01
%
%wfillfactor:
%minB6:
%running final lnlike to set model to best L  values
%chi2 = 1.241054e+00
%peakB6:
%running final lnlike to set model to best L  values
%chi2 = 1.133447e-01

% rsync -va   ~/common/ppdisks/ISO-Oph_2/mfreqfits/output_imoptim_wb3r07_wfillfactor/triangle_all.png  ./figs/triangle_all_wfillfactor.png 
% rsync -va   ~/common/ppdisks/ISO-Oph_2/mfreqfits/output_imoptim_wb3r07_wfillfactor/fig_bestfit_peakB6.png  ./figs/fig_bestfit_peakB6_wfillfactor.png
% rsync -va   ~/common/ppdisks/ISO-Oph_2/mfreqfits/output_imoptim_wb3r07_wfillfactor/fig_bestfit_minB6.png  ./figs/fig_bestfit_minB6_wfillfactor.png

\subsection{Fixing  $a_{\rm max}$}

\begin{figure}
  \centering
  \includegraphics[width=\columnwidth,height=!]{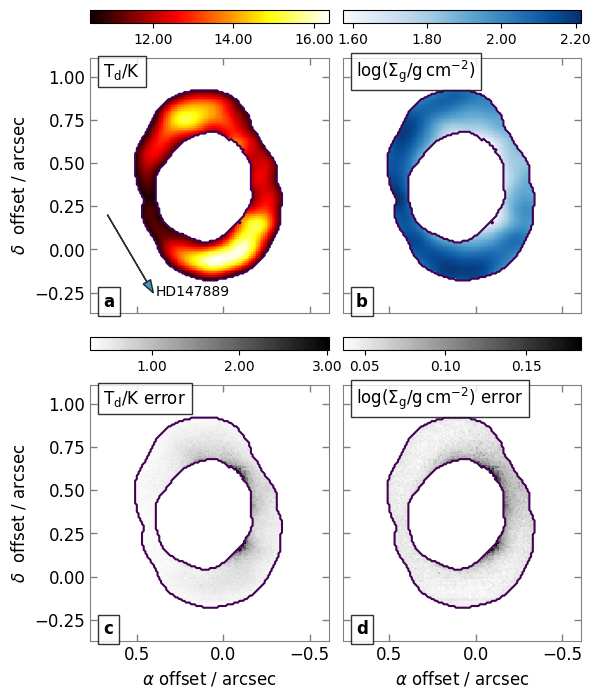}
  \caption{Dust parameters for each line of sight from the
    uniform-slab optimizations of $T_{\rm d}$ and $\Sigma_g$,
    constrained with the B8 and B6 data, degraded to the B8 beam with
    $r=0$. The black contours outline an intensity mask, set to
    $20\,\sigma$ in B8. {\bf a}: Maximum-likelihood $T_{\rm d}$. The
    arrow points to the direction of HD\,147889. {\bf b}:
    Maximum-likelihood $\log_{10}(\Sigma_g)$.  {\bf c}: One-sigma
    uncertainty on $T_{\rm d})$, approximated as
    $\sigma(T_{\rm d}) = \ln(10) T_{\rm d} \sigma(\log(T_{\rm
      d}))$. {\bf d}: One-sigma uncertainty on $\log_{10}(\Sigma_g)$.
  } \label{fig:imoptim_b6_b8}
\end{figure}

In Fig.\,\ref{fig:imoptim_b6_b8} we explore a fit with only two
free-parameters, i.e. $T_{\rm d}$ and $\Sigma_g$. In order to reach
the optically thick limit close to B8, and thus lift the
$T_{\rm d} -\Sigma_g$ degeneracy, we set $a_{\rm max} =
0.01\,$cm. This choice yields well-constrained posteriors for
$T_{\rm d}$ and $\Sigma_g$.  Fixing $a_{\rm max} \gtrsim 1\,$cm
results in optically thin emission to the West, and large errors on
$T_{\rm d}$.  However, in the East the morphologies of $T_{\rm d}$ and
$\Sigma_g$ are very similar in both cases.

%%(base) simon@uaren:~/common/ppdisks/ISO-Oph_2/report_mfreq$ rsync -va ~/common/ppdisks/ISO-Oph_2/mfreqfits/output_imoptim_b8r0_amax0.01_singlelosminB4/triangle_all.png   ./figs/triangle_b8r0.png
%
%\begin{figure}
%  \centering
%  \includegraphics[width=\columnwidth,height=!]{figs/triangle_b8r0.png}
%  \caption{Same as Fig.\,\ref{fig:corner} but for the fits to the B6
%    and B8 data only, and with two free-parameters, $T_{\rm d}$ and
%    $\Sigma_g$. Here we have set $a_{\rm max}= 0.01$\,cm. The
%    line-of-sight points towards the minimum in B4 along the ring,  at
%    $ \Delta \alpha = -0\farcs19$  $\Delta \delta = 0\farcs59$.  } \label{fig:cornerb8r0}
%\end{figure}

%~/common/ppdisks/ISO-Oph_2/mfreqfits$ python drive_imoptim_b8r0.py

%(base) simon@uaren:~/common/ppdisks/ISO-Oph_2/report_mfreq$ rsync -va ~/common/ppdisks/ISO-Oph_2/mfreqfits/output_imoptim_b8r0_amax0.01/fig_dustparams.png ./figs/fig_dustparams_b8r0_amax01.png
% ~/common/ppdisks/ISO-Oph_2/mfreqfits$ python gen_summary_dustparams.py

% rsync -va  ../mfreqfits/output_imoptim_b8r0_amax0.01_fine/fig_dustparams.png  ./figs/fig_dustparams_b8r0_amax01_fine.png
% rsync -va  ../mfreqfits/output_imoptim_b8r0_amax0.01_fine/fig_dustparams_linear.png  ./figs/fig_dustparams_b8r0_amax01_fine_linear.png

\subsection{Discussion}

\begin{figure}
  \centering
  \includegraphics[width=\columnwidth,height=!]{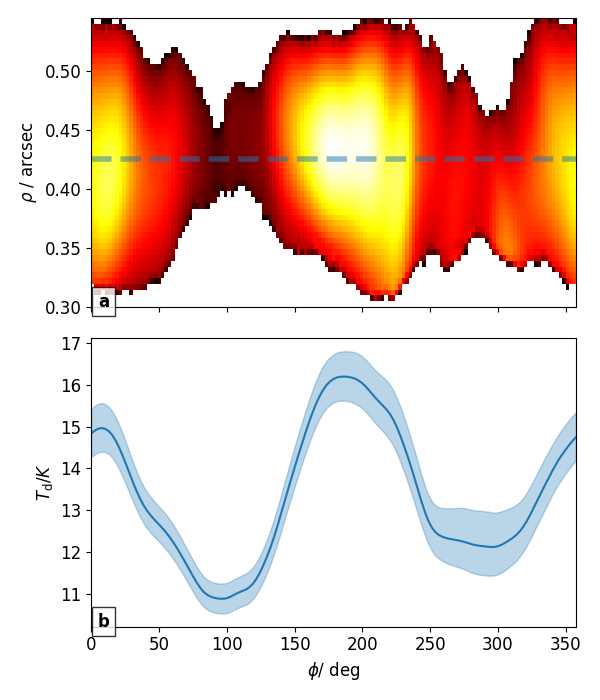}
  \caption{Polar expansion of the temperature map from
    Fig.\,\ref{fig:imoptim_b6_b8}a, shown with the same colour scale,
    and its azimuthal profile, as extracted at a radius
    $\rho=0\farcs425$. The shaded area corresponds to the region
    enclosed by the
    1-$\sigma$ uncertainties (its total vertical extent is
    2$\sigma$). The extraction radius is indicated as a dashed line in
    $a)$.  } \label{fig:az}
\end{figure}

An interesting result of the present estimates of physical conditions
are the significant variations in $T_{\rm d}$ along the ring, and
especially along the eastern arc. In the fits including B3, the
minimum along the ring is $\log(T_{\rm d}/K)=1.04\pm0.04$, while
$T_{\rm d}$ reaches $\log(T_{\rm d}/K)=1.32\pm0.08$ to the North and
South. The fits to the B6 and B8 data reach higher $T_{\rm d}$, as
expected since beam-dilution is reduced, and cover all azimuth. In
Fig.\,\ref{fig:az} we extracted the azimuthal profile for $T_{\rm d}$,
for which we adopted the orientation from the B8 estimates,
i.e. PA=7.8\,deg and $i=36.6$\,deg. The variations in $T_{\rm d}$ are
quite significant, and reach $\log(T_{\rm d}/K)= 1.175\pm0.017$ in the
North, and $\log(T_{\rm d}/K)= 1.210\pm0.015$ in the South, with a
minimum towards the East at $\log(T_{\rm d}/K)= 1.036\pm0.014$, which
represents over $8\sigma$.

%peaknorth 1.1751995186501447  +  0.01735344977114683  -  0.016433104784879526
%Tmin 1.0363655595470287  +  0.014564177412360674  -  0.014152792548371948
%phimax1  7.76978417266187
%phimax2  186.47482014388487
%peaksouth 1.2095859618580491
%peaksouth 1.036560008895575  +  0.014612847347479823  -  0.01411440197372272

Interestingly, the PA on the sky of the line joining the two peaks in
Fig.\,\ref{fig:az}b is 193.5\,deg, and is remarkably close to the
direction of HD\,147889, which is at 210.5\,deg.  The azimuthal
temperature modulation might thus result from a variation of the angle
of incidence of radiation coming from HD\,147889. Such external
irradiation would hit the Southern edge of the disk almost edge-on,
and, since the disk is flared, the region where it would reach the
disk surface at closest to normal incidence is to the North.  The
small difference between the PA joining the two temperature maxima
might be due to biases in our estimate of the dust temperature, since
here we kept the dust grain size fixed at a small value that ensures
that B8 is in the optically thick regime. Another interesting
possibility is that, if the disk is retrograde (rotating clock-wise),
then the small angular shift could be due to the thermal lag discussed
in \citet{Casassus2019MNRAS.486L..58C}.

%PApeaks -166.47161964673842
  
%(base) simon@uaren:~/common/ppdisks/ISO-Oph_2/report_mfreq$ rsync -va ~/common/ppdisks/ISO-Oph_2/mfreqfits/polarmaps/polarmaps_logTd/fig_prof_Td.png ./figs/
%~/common/ppdisks/ISO-Oph_2/mfreqfits/polarmaps/polarmaps_logTd/Mextract_profile.py
% rsync  -va  ../mfreqfits/polarmaps/polarmaps_logTd/fig_prof_Td_linear.png ./figs/ 

\section{Conclusions} \label{sec:conc}

We report new ALMA continuum observations of the ISO-Oph\,2 binary, at 97\,GHz, 145\,GHz and 405\,GHz, that complement existing 225\,GHz data. A novel strategy for the alignment of multi-frequency data, acquired with broadly different angular resolutions, allowed us to reach the following conclusions:
\begin{enumerate}
\item The offset of ISO-Oph\,2A relative to the centroid of the
  circumprimary disk is remarkably large, of 62\,mas to 76\,mas
  depending on the image (Fig.\,\ref{fig:summary_mfreq}). Such a large
  offset points at dynamical interactions, either with ISO-Oph\,2B or
  with other massive bodies inside the ring of ISO-Oph\,2A.
\item The multi-frequency data reveal strong morphological variations
  with frequency in   ISO-Oph\,2A (Fig.\,\ref{fig:rgb}). We linked these variations
  to the underlying physical conditions by modeling the data with
  uniform-slab intensities (Figs.\,\ref{fig:imoptim_b3_b6_b8} and
  \ref{fig:imoptim_b6_b8}).
\item Surprisingly, the dust temperature varies strongly in azimuth
  (Fig.\,\ref{fig:az}), and roughly traces a second harmonic with 4
  nodes. The PA joining the two peaks, each to the North and South of
  the disk, is aligned in the direction towards HD\,147889 within
  10\,deg. Such an azimuthal temperature modulation is in qualitative
  agreement with external irradiation as the dominant heat source.
\item As in several other disks, we find indications for a lopsided disk, where the dust column density is shaped into  a crescent. The maximum grain size appears to coincide with the peak column density, as expected for aerodynamic dust trapping (Fig.\,\ref{fig:imoptim_b3_b6_b8}). 
\item The multi-epoch astrometry of the  binary
  is only marginally consistent with a bound orbit, in support (but at $\sim 2\sigma$)
of the view that the binary is in fact a fly-by.
\end{enumerate}

The binary disks of ISO-Oph\,2 are interesting laboratories for the
impact of environmental effects on disk structure, with strong
dynamical perturbations on the circumprimary ring. The temperature
structure of this ring is also suggestive of heating by external
irradiation, probably from HD\,147889.  This possibility will be
considered in a companion article on radiative transfer modeling of
external irradiation in ISO-Oph\,2.

\section*{Acknowledgements}

We thank the two referees (an anonymous referee and Prof. Takayuki
Muto) for their constructive comments. S.C., L.C. and M.C. acknowledge
support from Agencia Nacional de Investigaci\'on y Desarrollo de Chile
(ANID) given by FONDECYT Regular grants 1211496, 1211656, ANID
PFCHA/DOCTORADO BECAS CHILE/2018-72190574, ANID project Data
Observatory Foundation DO210001, and ANID - Millennium Science
Initiative Program - Center Code NCN2021 080. A.R. has been supported
by the UK Science and Technology research Council (STFC) via the
consolidated grant ST/W000997/1 and by the European Union’s Horizon
2020 research and innovation programme under the Marie
Sklodowska-Curie grant agreement No. 823823 (RISE DUSTBUSTERS
project).  VC acknowledges a postdoctoral fellowship from the Belgian
F.R.S.-FNRS.  T.B. acknowledges financial support from the FONDECYT
postdoctorado project number 3230470. A.R.-J. acknowledge funding from
ANID-Subdirección de Capital Humano/Doctorado Nacional/2022-21221841.
This paper makes use of the following ALMA data: {\tt
  ADS/JAO.ALMA\#2022.1.01734.S}, {\tt \#2021.1.00378.S}, {\tt
  2019.1.01111.S} and {\tt \#2018.1.00028.S}. ALMA is a partnership of
ESO (representing its member states), NSF (USA) and NINS (Japan),
together with NRC (Canada), MOST and ASIAA (Taiwan), and KASI
(Republic of Korea), in cooperation with the Republic of Chile. The
Joint ALMA Observatory is operated by ESO, AUI/NRAO and NAOJ.

%%%%%%%%%%%%%%%%%%%%%%%%%%%%%%%%%%%%%%%%%%%%%%%%%% 
\section*{Data Availability}
The reduced ALMA data presented in this article are available upon
reasonable request to the corresponding author.
The original or else non-standard software packages 
underlying the analysis are available at the following URLs: {\sc
  MPolarMaps}
\citep[\url{https://github.com/simoncasassus/MPolarMaps},][]{Casassus2021MNRAS.507.3789C}, {\sc uvmem}
\citep[\url{https://github.com/miguelcarcamov/gpuvmem},][]{Carcamo2018A&C....22...16C}, {\sc pyralysis}
(\url{https://gitlab.com/clirai/pyralysis}), {\sc VisAlign}
(\url{https://github.com/simoncasassus/VisAlign}), {\sc snow}
(\url{https://github.com/miguelcarcamov/snow}).



\bibliographystyle{mnras}
%\bibliography{mfreq} % if your bibtex file is called example.bib

\input{mfreq_ISO-Oph_2.bbl}

%\begin{thebibliography}{99}
%\end{thebibliography}

%%%%%%%%%%%%%%%%%%%%%%%%%%%%%%%%%%%%%%%%%%%%%%%%%%

%%%%%%%%%%%%%%%%% APPENDICES %%%%%%%%%%%%%%%%%%%%%

\appendix

\section{On the JvM correction} \label{sec:JvM}

The so-called ``JvM correction'' \citep[][]{JvM1995AJ....110.2037J,
  Czekala2021ApJS..257....2C} is thought to improve the dynamic range
of images restored from radio-interferometric data. However, here we
did not apply the JvM correction, because the resulting improvement is
due to a spurious down-scaling of the image residuals, as shown in
\citet[][]{CasassusCarcamo2022MNRAS.513.5790C}. Despite the proof,
since its publication several workers have kept on applying the JvM
correction, which leads us to believe that perhaps the arguments
presented in \citet[][]{CasassusCarcamo2022MNRAS.513.5790C} may 
not be clear enough. Here we  give more details on the argumentation that defines the units of the dirty map in interferometric image reconstruction. 

As summarised in Appendix\,A of \citet[][Eq.\,A2
]{CasassusCarcamo2022MNRAS.513.5790C}, the restored image  is
obtained by adding the dirty map $R_D$ of the visibility residuals
with the model image $I_m$, after convolution with the clean beam
$g_b$:
\begin{equation}
  I_R = I_m \ast g_b + R_D .  \label{eq:IR}
\end{equation}
Both the convolved model image and dirty residuals must of course bear
the same units. \citet{CasassusCarcamo2022MNRAS.513.5790C} proposed to
tie these units to the case of a point source at the phase center,
where the flux of the point source and its uncertainty can be inferred
from parametric modelling of the visibility data (e.g. their Eqs.\,A11
and A12). They matched this uncertainty to the thermal uncertainty on
the specific intensity in the dirty map at the phase center (their
Eqs.\,A13 and A14).

However,  \citet{CasassusCarcamo2022MNRAS.513.5790C} did not explain the relationship
between the general expression for the dirty map $I_D$ (originally in
Eq.\,A9) and that of the residuals $R_D$ in Eq.\,\ref{eq:IR}
above. Here we clarify that, for the test-case of a point source at
the phase center, the uncertainties on $I_R$ stem from the thermal
noise in $R_D$, since the model of the source is known. The dirty map
$R_D$ is itself an application of the general formula for $I_D$ to the
residual visibilities of the parametric fit. These residuals should
contain only noise in this idealised test-case.

\section{Brightness temperature maps} \label{sec:Tbmaps}

\begin{figure*}
  \centering
  \includegraphics[width=0.8\textwidth,height=!]{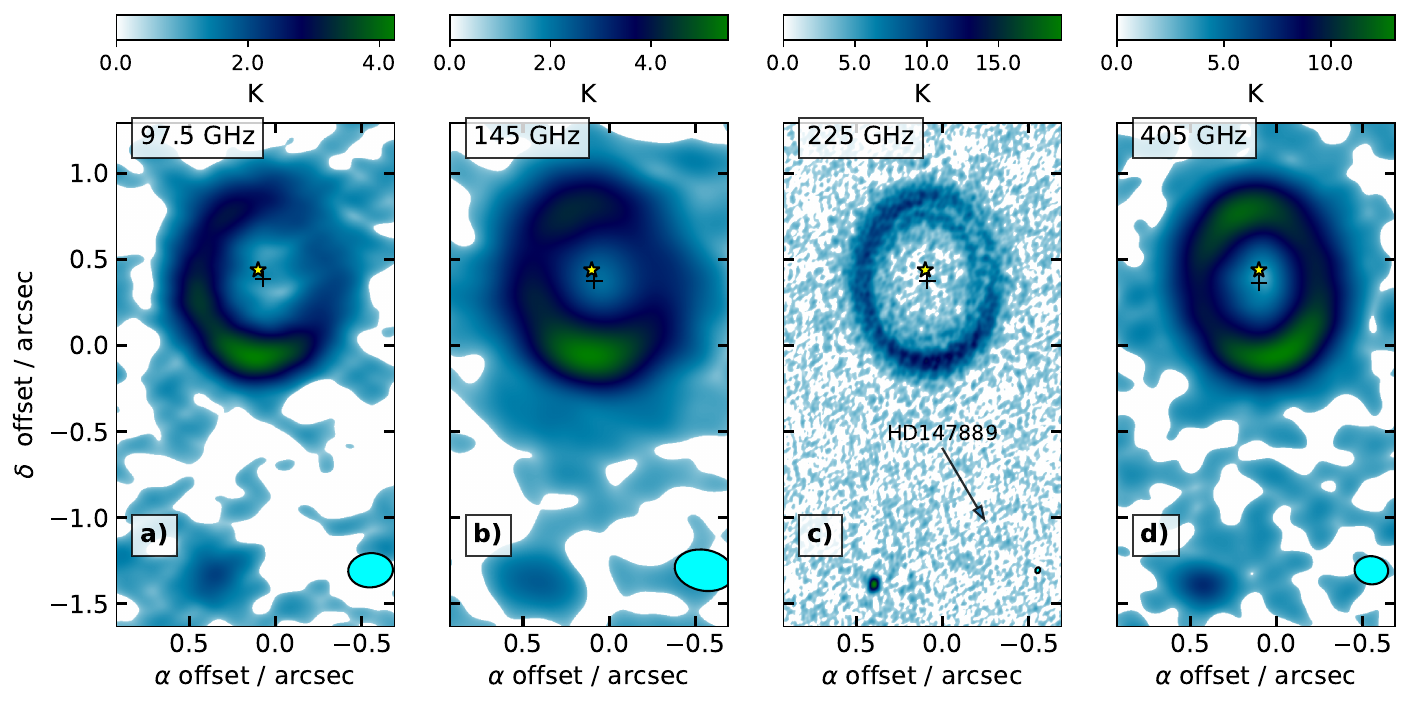}
  \caption{Same as Fig.\,\ref{fig:summary_mfreq}, but for brightness
    temperature maps. The images have been clipped at
    0\,K.} \label{fig:summary_mfreq_Tb}
\end{figure*}

Fig.\,\ref{fig:summary_mfreq_Tb} includes a summary of the
self-calibrated and aligned data, as in Fig.\,\ref{fig:summary_mfreq},
but in brightness temperature.

%\section{Alignment of  visibility datasets}

\section{Figure of merit for the alignment of  visibility datasets} \label{sec:figofmerit}

The alignment of the multi-frequency visibility data  was
performed with the {\sc
  VisAlign}\footnote{\url{https://github.com/simoncasassus/VisAlign}}
package, described in
\citet[][]{CasassusCarcamo2022MNRAS.513.5790C}. However, here we
improved {\sc VisAlign} with an adjustment to the figure of merit, as
the least-squares formula associated to the alignment of two
visibility datasets, i.e. Equation\,1 in
\citet[][]{CasassusCarcamo2022MNRAS.513.5790C}, was biased in the
choice of reference dataset. In this appendix we revisit the least-squares figure of merit used to  perform the alignment of two visibility datasets, i.e. Equation\,1 in \citet[][]{CasassusCarcamo2022MNRAS.513.5790C}, which we  reproduce here for clarity:
\begin{equation}
  \chi^2_{\rm align}(\alpha_R, \delta{\vec{x}}) = \sum_{k=1}^N W^{\rm align}_k \| \tilde{V}_k^L - \tilde{V}_k^{Lm}   \|^2, \label{eq:chi2}
\end{equation}
where
\begin{equation}
  \tilde{V}_k^{Lm} = \alpha_R \, e^{i 2\pi \,\delta\vec{x}\cdot\vec{u}_k} \, \tilde{V}_k^S,
\end{equation}
and 
\begin{equation}
  W_k^{\rm align} = \frac{W_k^S \, W_k^L}{W_k^S+  W_k^L}. \label{eq:commonW}
\end{equation}
With such  weights $W_k$,  the minimization of $\chi^2_{\rm align}$ in Eq.\,\ref{eq:chi2} is not
symmetric in the choice of reference dataset for the alignment of the two visibilty datasets
$\left\{\tilde{V}^S_k\right\}_{k=1}^N$ and
$\left\{\tilde{V}^L_k \right\}_{k=1}^N $. In other words, aligning
$\tilde{V}^S$ to $\tilde{V}^L$ does not yield the opposite shift and
reciprocal flux correction as aligning $\tilde{V}^L$ to $\tilde{V}^S$.
A symmetric expression, now implemented in the {\sc
  VisAlign}\footnote{\url{https://github.com/simoncasassus/VisAlign}}
package, is obtained by replacing the weights with:
\begin{equation}
  W_k^{\rm align} = \frac{W_k^S \, W_k^L}{W_k^S+  \alpha_R^2 W_k^L}. \label{eq:commonWfixed}
\end{equation}
We confirmed that with this modification the alignment is now
independent on the choice of reference data-set, in the sense that the
astrometric shifts are opposite and the flux scale factors are
reciprocal (down to the round-off numerical accuracy).

The impact on the  corresponding flux scale factors and
astrometric shifts is small ($\sim 5\%-10\%$). For example, following
the nomenclature of \citet{Benisty2021ApJ...916L...2B} for the each
visibility dataset, the updated flux scale correction factors are
$\alpha_R = 0.817\pm0.003$ to match SB16 to LB19, and
$\alpha_R = 0.837\pm0.002$ to match IB17 and LB19.

Another consequence is that the shifts are no
longer sensitive on the choice of $uv-$range, and depend only on the
choice of $uv$-plane cell size for gridding, $\Delta u$. We checked
that the shifts are all consistent within the errors for widely
different choices of $\Delta u$, ranging from the antenna diameter to
the minimum baseline length.

\section{Update to the multi-epoch radio-continuum imaging of PDS\,70} \label{sec:pds70}

The impact of the revised alignment on the corresponding flux scale
factors and astrometric shifts, although  small ($\sim 5\%-10\%$),  affects the multi-epoch analysis   of PDS\,70 reported in \citet[][]{CasassusCarcamo2022MNRAS.513.5790C}. Here we update the resulting images. All the conclusions from \citet[][]{CasassusCarcamo2022MNRAS.513.5790C} hold, but the variability of PDS\,70c is more significant.

As in \citet[][]{CasassusCarcamo2022MNRAS.513.5790C}, we self-calibrated each data-set
individually before concatenation. Self-calibration was performed
automatically with the {\sc OOselfcal} package, which we re-baptised
to ``Self-calibratioN Object-oriented frameWork'', i.e. {\sc snow}\footnote{see Data Availability}. A consequence of the updated figure
of merit is that the peak signal-to-noise ratios for the concatenated
datasets are already close to the values obtained after joint
self-calibration.

The LB19 dataset was used as reference for the alignment of the
multi-epoch data. However, the 2020 {\em GAIA} coordinates
\citep[DR\,3,][]{2020yCat.1350....0G} for PDS\,70 are offset by
9.2\,mas relative to the LB19 phase center, by 8.2\,mas in R.A. and
-4.2\,mas in Dec.. This shift is larger than the nominal pointing
accuracy of the LB19 dataset (whose standard deviation  is about a tenth of a beam or
$\sim$5\,mas).

\begin{figure*}
  \centering
  \includegraphics[width=\textwidth,height=!]{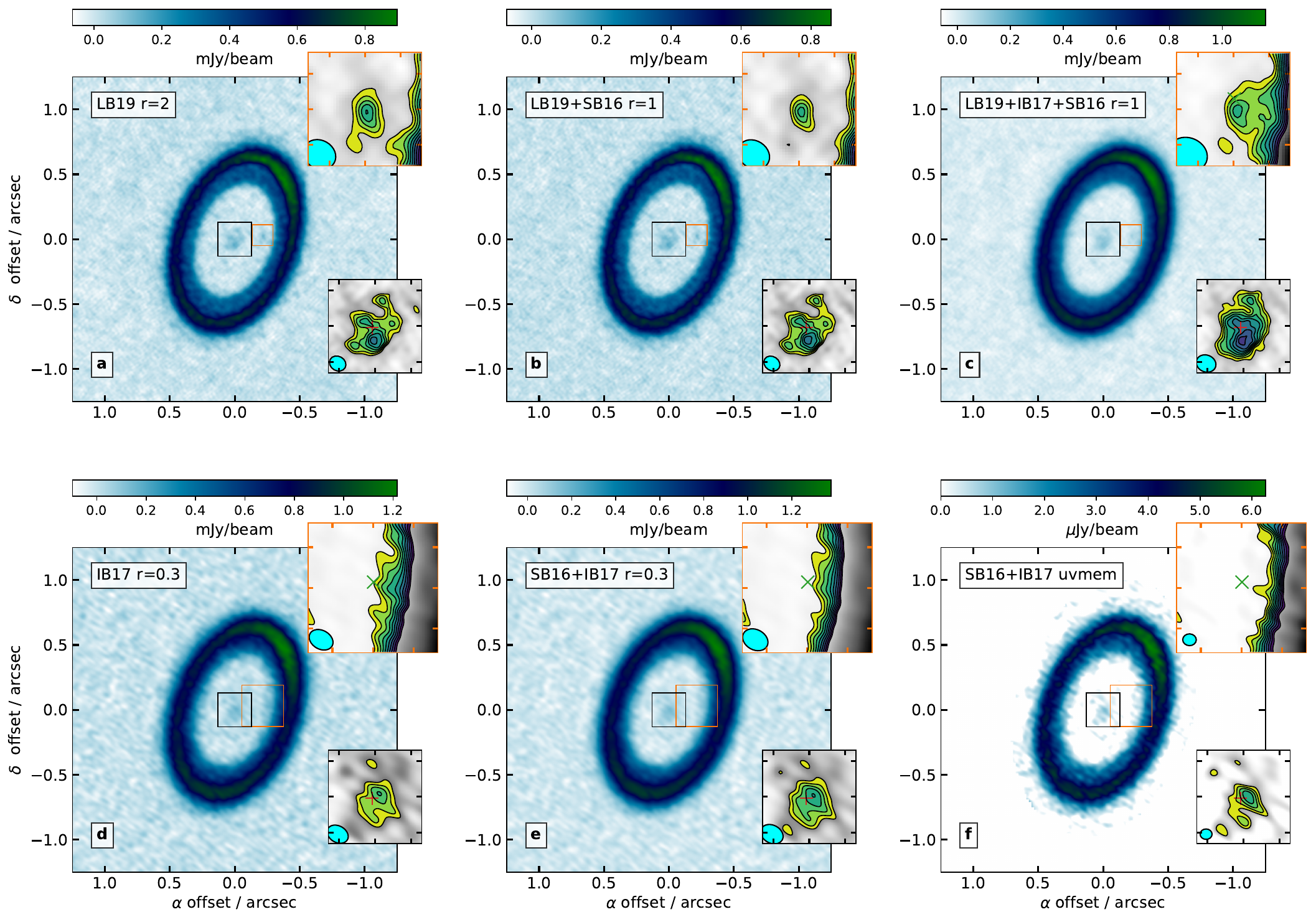}
  \caption{
    Annotations follow from Fig.\,4 in
    \citet[][]{CasassusCarcamo2022MNRAS.513.5790C}, except for the red
    cross in the insets on the inner disk, which is now pointed at the
    latest {\em GAIA} coordinates for PDS\,70. The updated beam and noise values follow.
    {\bf a:} Clean beam    
    $\Omega_b = 0\farcs048\times 0\farcs040 \,/\,61$\,deg. The noise in the
    residual image is $\sigma = 16.6\,\mu$Jy\,beam$^{-1}$.
    {\bf b:}
    $\Omega_b = 0\farcs047\times0\farcs039\,/\,57$deg,
    $\sigma = 15.5\,\mu$Jy\,beam$^{-1}$.
    {\bf c:} 
    $\Omega_b = 0\farcs056\times0\farcs047\,/\,67$deg, and
    $\sigma = 13.1\,\mu$Jy\,beam$^{-1}$.
    {\bf d:}  
    $\Omega_b = 0\farcs062\times 0\farcs046 \,/\,60$\,deg, and
    $\sigma = 25.9\,\mu$Jy\,beam$^{-1}$.  
    {\bf e:} 
    $\Omega_b = 0\farcs066\times 0\farcs050 \,/\,61$\,deg, and
    $\sigma = 23.8\,\mu$Jy\,beam$^{-1}$.
    {\bf f:} approximate resolution of 1/3 the
    natural weight beam \citep[][]{Carcamo2018A&C....22...16C}, or
    $\Omega_b \approx 0\farcs033\times 0\farcs029 \,/\,88$\,deg. The
    contours start at 3\,$\sigma$, where
    $\sigma = 0.16\,\mu$Jy\,pix$^{-1}$
    is a representative noise level (with 4\,mas pixels).} \label{fig:PDS70_concats_all}
\end{figure*}

In addition to the correction on the alignment procedure,
the scheduling block from Dec. 6, 2017, was missing in the images for
the IB17 dataset re-processed in
\citet[][]{CasassusCarcamo2022MNRAS.513.5790C}, who therefore included  
 only 2/3 of the available dataset. The incorporation of this
scheduling block improves the sensitivity of the IB17 images, and
results in tighter constraints on the absence of PDS\,70c in the IB17
data.  The corrected images are shown in
Fig.\,\ref{fig:PDS70_concats_all}.

A final correction to the analysis presented in \citep[][]{Carcamo2018A&C....22...16C} concerns the choice of reference frequency for multi-frequency synthesis.
In the {\sc uvmem} imaging package
\citep[][]{Carcamo2018A&C....22...16C}, multi-frequency synthesis is
implemented with two options. The user can select to fit a spectral
index map $\alpha(\vec{x})$ to the data, or use a single and constant
spectral index value $\alpha$ to propagate the model visibilities to
all frequencies (in specific intensity units, i.e
$I_\nu = I_\circ (\nu / \nu_\circ)^\alpha$). We usually adopt a flat
spectral index, $\alpha = 0$, but 
\citet{CasassusCarcamo2022MNRAS.513.5790C}  opted to fix
$\alpha =3$, with a reference frequency taken as the median of the
centroid frequencies of all spectral windows in the concatenated
datasets. This choice of reference frequency is slightly different
from the default in CASA {\tt tclean}, which uses the middle
frequency. We have now unified the choice of frequency, as required
for  image restoration. With this correction the point source in
LB19 coincident with PDS\,70c is now also visible in the concatenation
LB19+IB17+SB16 (see Fig.\,\ref{fig:PDS70_concats_all}c).

%For the alignment of SB16 to LB19, the
%corrected values are
%$\delta\vec{x} = (-0\farcs011 \pm 0\farcs0004, -0\farcs017\pm0\farcs
%0005)$ and $\alpha_R = 0.817\pm0.003$. For the alignment of
%IB17 to LB19, the flux scale factor is $\alpha_R = 0.958\pm0.003$, and the 

\begin{figure}
  \begin{center}
  \includegraphics[width=\columnwidth,height=!]{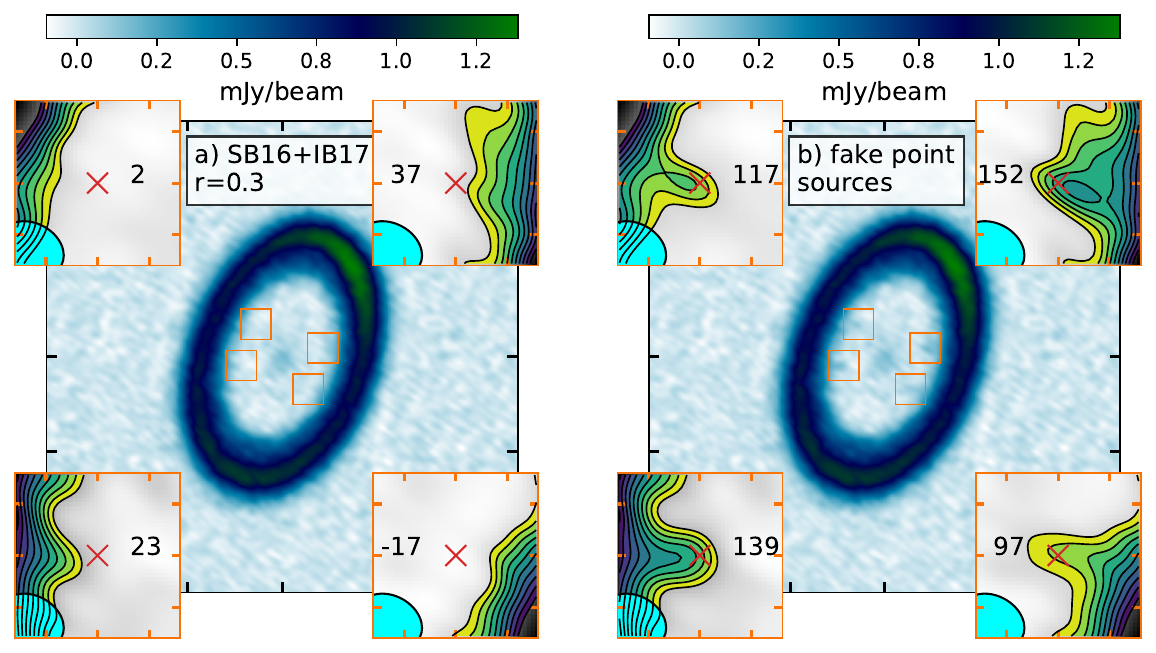}
  \caption{Annotations follow from Fig.\,6 in \citet[][]{CasassusCarcamo2022MNRAS.513.5790C}. \label{fig:PSs} }
  \end{center}
\end{figure}

The noise level in the residuals for the SB16+IB17 image is
23.9\,$\mu$Jy\,beam$^{-1}$ (versus 31.9\,$\mu$Jy\,beam$^{-1}$ in our
original publication). The point source coincident with PDS\,70c in
the LB19 dataset, with peak flux 118.5\,$\mu$Jy\,beam$^{-1}$ (from
Fig.\,\ref{fig:PDS70_concats_all}a), should have been picked up in
IB17 at 5.0$\sigma$. The point source injections tests are accordingly
updated in Fig.\,\ref{fig:PSs}. With these new numbers, if we assign a
3$\sigma$ upper limit flux for PDS\,70c in IB17, then it was fainter
by $40\%\pm8\%$ relative to LB19 (versus $42\% \pm 13\% $ in the
original publication).

\begin{figure}
  \centering
  \includegraphics[width=\columnwidth,height=!]{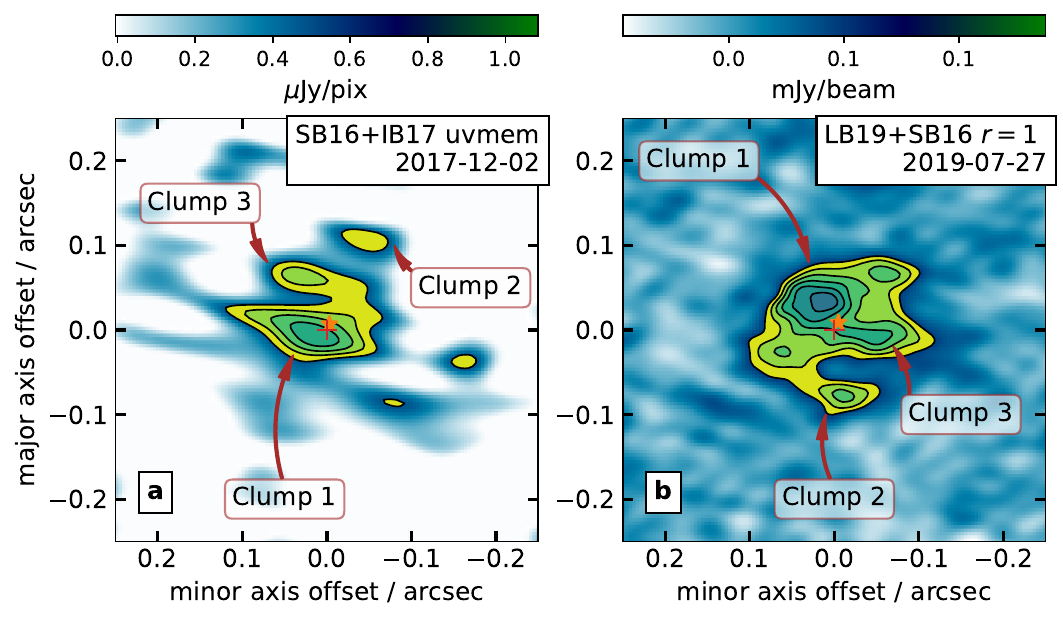}
  \caption{Annotations follow from Fig.\,7 in \citet[][]{CasassusCarcamo2022MNRAS.513.5790C}, except that the red plus sign is now centered on the ring centroid (at the origin of coordinates), while the orange star indicates the position of the star. } \label{fig:PDS70_cav}
\end{figure}

An update on the face-on views of the inner disk, and its variability,
is given in Fig.\,\ref{fig:PDS70_cav}, including the updated stellar
position. The relative pointing accuracy of the multi-epoch data, as
estimated from {\sc VisAlign}, is $\sim 0.4$\,mas, but the absolute
pointing accuracy of the LB19 data-set  is $\sim$5\,mas and
affects both epochs equally (in the same direction). The accuracy on
the position of the ring centroid is 0.5\,mas in SB16+LB19 and
0.7\,mas in SB16+IB17. The offset between the nominal stellar position
and the ring centroid is 8\,mas in IB17, and 10\,mas in LB19.

In summary, the improvements to the analysis of the multi-epoch
radio-continuum data from PDS\,70 lead to tigher constraints on the
variability of PDS\,70c. The associated point source is variable by at
least 40$\pm$ 8\% in 1.75\,yr, assigning the upper limit flux of
3$\sigma$ in the 2017.

%which together with other improvements leads to
%tigher constraints on the variability of PDS\,70c, of at least 40$\pm$
%8\% in 1.75\,yr, assigning the upper limit flux of 3$\sigma$ in the
%2017.
%\item The multi-epoch imaging was tied to the pointing center and flux scale of one of datasets, the LB19 dataset in the nomenclature of \citet[][]{Benisty2021ApJ...916L...2B}, whose phase center was slightly offset from the 2020 {\em GAIA} coordinates for PDS\,70.
%\item We incorporate  a missing scheduling block in the IB17 observations of PDS\,70 \citep[originally presented in][]{Isella2019ApJ...879L..25I}, that leads to improved imaging.
%\item We unify the reference frequencies for multi-frequency imaging when concatenating visibility data.
%\item We give more details on the argumentation that defines the units of the dirty map in interferometric image reconstruction. 
%\
%

%%%%%%%%%%%%%%%%%%%%%%%%%%%%%%%%%%%%%%%%%%%%%%%%%%

% Don't change these lines
\bsp	% typesetting comment
\label{lastpage}
\end{document}